\documentclass[fleqn,usenatbib]{mnras}

\usepackage{newtxtext,newtxmath}
\usepackage{siunitx}    % units
\usepackage{multirow}

\usepackage[T1]{fontenc}

\DeclareRobustCommand{\VAN}[3]{#2}
\let\VANthebibliography\thebibliography
\def\thebibliography{\DeclareRobustCommand{\VAN}[3]{##3}\VANthebibliography}

\usepackage{graphicx}	% Including figure files
\usepackage{amsmath}	% Advanced maths commands
\newcommand{\guvb}{\Gamma_{\text{HI}}}
\newcommand{\lya}{Ly$\alpha$\ }
\newcommand{\lyb}{Ly$\beta$\ }
\newcommand{\lammfp}{\lambda_{\text{mfp}}}
\newcommand{\lamobs}{\lambda_{\text{obs}}}
\newcommand{\lamrest}{\lambda_{\text{rest}}}
\newcommand{\xidif}{\frac{\xi_F - \langle F \rangle^2}{\langle F \rangle^2}}
\title[\lya flux auto-correlation with XQR-30]{Measurements of the $z > 5$ Lyman-$\alpha$ forest flux auto-correlation functions from the extended XQR-30 data set}

\author[Molly Wolfson et al.]{Molly Wolfson,$^{1}$\thanks{E-mail: mawolfson@ucsb.edu}
Joseph F. Hennawi$^{1,2}$,
Sarah E. I. Bosman$^{3,4}$,
Frederick B. Davies$^{3}$, \newauthor
Zarija Luki\'{c}$^{5}$,
George D. Becker$^{6}$, 
Huanqing Chen$^{7}$,
Guido Cupani$^{8, 9}$,
Valentina D'Odorico$^{8,9,10}$, \newauthor
Anna-Christina Eilers$^{11}$, 
Martin G. Haehnelt$^{12}$,
Laura C. Keating$^{13}$,
Girish Kulkarni$^{14}$,
Samuel Lai$^{15}$, \newauthor
Andrei Mesinger$^{10}$, 
Fabian Walter$^{3}$, 
and Yongda Zhu$^{6}$
\\
$^{1}$Department of Physics, University of California, Santa Barbara, CA 93106, USA\\
$^{2}$Leiden Observatory, Leiden University, Niels Bohrweg 2, 2333 CA Leiden, Netherlands\\
$^{3}$Max-Planck-Institut f\"{u}r Astronomie, K\"{o}nigstuhl 17, 69117 Heidelberg, Germany\\
$^{4}$Institute for Theoretical Physics, Heidelberg University, Philosophenweg 12, D-69120, Heidelberg, Germany\\
$^{5}$Lawrence Berkeley National Laboratory, 1 Cyclotron Rd, Berkeley, CA 94720, USA\\
$^{6}$Department of Physics \& Astronomy, University of California, Riverside, CA 92521, USA\\
$^{7}$Canadian Institute for Theoretical Astrophysics, University of Toronto, 60 St George St, Toronto, ON. M5S 3H8, Canada\\
$^{8}$INAF-Osservatorio Astronomico di Trieste, Via Tiepolo 11, I-34143 Trieste, Italy\\
$^{9}$IFPU-Institute for Fundamental Physics of the Universe, via Beirut 2, I-34151 Trieste, Italy\\
$^{10}$Scuola Normale Superiore, Piazza dei Cavalieri 7, I-56126 Pisa, Italy\\
$^{11}$MIT Kavli Institute for Astrophysics and Space Research, 77 Massachusetts Avenue, Cambridge, MA 02139, USA\\
$^{12}$Kavli Institute for Cosmology and Institute of Astronomy, Madingley Road, Cambridge, CB3 0HA, UK\\
$^{13}$Institute for Astronomy, University of Edinburgh, Blackford Hill, Edinburgh, EH9 3HJ, UK\\
$^{14}$Tata Institute of Fundamental Research, Homi Bhabha Road, Mumbai 400005, India\\
$^{14}$Research School of Astronomy and Astrophysics, Australian National University, Canberra, ACT 2611, Australia\\
}

\date{Accepted XXX. Received YYY; in original form ZZZ}

\pubyear{2023}

\begin{document}
\label{firstpage}
\pagerange{\pageref{firstpage}--\pageref{lastpage}}
\maketitle

\begin{abstract}
Recently, the Lyman-$\alpha$ (Ly$\alpha$) forest flux auto-correlation function has been shown to be sensitive to the mean free path of hydrogen-ionizing photons, $\lambda_{\text{mfp}}$, for simulations at $z \geq 5.4$. 
Measuring $\lambda_{\text{mfp}}$ at these redshifts will give vital information on the ending of reionization. 
Here we present the first observational measurements of the Ly$\alpha$ forest flux auto-correlation functions in ten redshift bins from $5.1 \leq z \leq 6.0$.
We use a sample of 35 quasar sightlines at $z > 5.7$ from the extended XQR-30 data set, this data has signal-to-noise ratios of $> 20$ per spectral pixel. 
We carefully account for systematic errors in continuum reconstruction, instrumentation, and contamination by damped Ly$\alpha$ systems. 
With these measurements, we introduce software tools to generate auto-correlation function measurements from any simulation. 
For an initial comparison, we show our auto-correlation measurements with simulation models for recently measured $\lambda_{\text{mfp}}$ values and find good agreements.
Further work in modeling and understanding the covariance matrices of the data is necessary to get robust measurements of $\lambda_{\text{mfp}}$ from this data. 
\end{abstract}


%%%%%%%%%%%%%%%%%%%%%%%%%%%%%%%%%%%%%%%%%%%%%%%%%%

%%%%%%%%%%%%%%%%% BODY OF PAPER %%%%%%%%%%%%%%%%%%

\section{Introduction}

The reionization of the neutral hydrogen in the intergalactic medium (IGM) is one of the major phase changes in our Universe's history. 
Understanding the timing of this process has been the focus of many recent studies. 
Current Planck constraints put the midpoint of reionization at $z_{\text{re}} = 7.7 \pm 0.7$ \citep{planck_2018} with mounting evidence that it was not completed until after $z \leq 6$ \citep{fan_2006, becker_2015, becker_2018, bosman_2018, bosman_2021_data, eilers_2018, boera_2019, yang_2020, jung_2020, kashino_2020, moreales_2021}. 

Before the end of reionization, the mean free path of hydrogen-ionizing photons ($\lammfp$) is expected to be short due to the significant neutral hydrogen remaining in the IGM which will absorb these photons close to their sources.
In some models, as reionization ends $\lammfp$ will rapidly increase due to the overlap of initially isolated ionized bubbles and the photo-evaporation of dense photon sinks \citep{gnedin_2000, shapiro_2004, furlanetto_2005, gnedin_fan_2006, wyithe_2008, sobacchi_2014, park_2016, kulkarni_2019, keating_2020_a, keating_2020_b, nasir_daloisio_2020, cain_2021, gnedin_2022}. 
Thus detecting an increase in $\lammfp$ will provide insights into the end of reionization.

\citet{becker_2021} reported the first direct measurement of $\lammfp$ at $z \sim 6$ from stacked quasar spectra.
% They found $\lammfp = 9.09^{+1.62}_{-1.28}$ proper Mpc at $z = 5.1$ and $\lammfp = 0.75^{+0.65}_{-0.45}$ proper Mpc at $z = 6$ from stacked quasar spectra. 
% This measurement has been updated and expanded to $z = 5.31$ and $z = 5.65$ in \citet{zhu_2023}. 
\citet{zhu_2023} updated this measurement and added two additional redshift bins at $z = 5.31$ and $z = 5.65$. 
They found that $\lammfp = 9.33^{+2.06}_{-1.80}$, $5.40^{+1.47}_{-1.40}$, $3.31^{+2.74}_{-1.34}$, and $0.81^{+0.73}_{-0.48}$ pMpc at $z = $5.08, 5.31, 5.65, and 5.93, respectively. 
\citet{becker_2021} and \citet{zhu_2023} expanded on previous measurements of $\lammfp$ at $z \leq 5.1$ \citep{prochaska_2009, fumagalli_2013, omeara_2013, worseck_2014}. 
The \citet{zhu_2023} measurement has $\lammfp$ rapidly increasing between $z = 6$ and $z = 5.1$, potentially signalling the end of reionization.
The values at $z \geq 5.3$ are significantly smaller than extrapolations from previous lower $z$ measurements \citep{worseck_2014} based on a fully ionized IGM.
In addition, the value at $z \sim 6$ may cause tension with measurements of the ionizing output from galaxies \citep{cain_2021, davies_2021}. 

Alternative methods to constrain $\lammfp$ are needed to check the measurements discussed above and to constrain the timing of reionization in finer redshift bins. 
One such method from \citet{bosman_2021_limit} used lower limits on individual free paths (the distance ionizing radiation travels from an individual source) towards high-$z$ sources to place a $2 \sigma$ limit of $\lammfp > 0.31$ proper Mpc at $z = 6.0$. 
This \citet{bosman_2021_limit} method is similar to other measurements using individual free paths \citep{songaila_2010, rudie_2013, romano_2019}.
Additionally, \citet{gaikwad_2023_mfp} constrained $\lammfp$ for $4.9 < z < 6.0$ with $\Delta z = 0.1$ by comparing the observed probability distribution function of the \lya optical depth to predictions from simulations with a fluctuating ultraviolet background (UVB) driven by a short $\lammfp$. 
The measurement of $\lammfp$ at $z < 5.1$ in \citet{gaikwad_2023_mfp} shows a good agreement with the measurements from \citet{worseck_2014} and \citet{becker_2021}. 
At $z = 6.0$ \citet{gaikwad_2023_mfp} measured $\lammfp = 8.318^{+7.531}_{-4.052}$ comoving Mpc (cMpc) h$^{-1}$, which agrees with the \citet{becker_2021} measurement at the $1.2 \sigma$ level and also falls above the lower limit found by \citet{bosman_2021_limit}. 

The level of fluctuations in the UVB, $\guvb$, are set by the distribution of ionizing photon sources and $\lammfp$. 
For large values of $\lammfp$, photons travel further from their sources and effectively creates a more uniform UVB \citep{mesinger_furlanetto_2009}.
Alternatively, small values of $\lammfp$ lead to greater fluctuations in the UVB, causing some regions to have very large $\guvb$ values. 
These fluctuations then imprint themselves on the \lya forest flux transmission in high-$z$ quasar spectra via the Ly$\alpha$ opacity, $\tau_{\rm Ly\alpha}$ where $\tau_{\rm Ly\alpha} = n_{\rm HI} \sigma_{{\rm Ly}\alpha} \propto  1 / \Gamma_{\rm HI} \propto 1 / \lammfp^{\alpha}$ where $3/2 < \alpha < 2$ \citep[see e.g.][]{Rauch_1998, haardt_madau_2012}. 
Many previous studies have investigated the effect of large scale variations in the UVB on the structure of the \lya forest \citep{zuo_1992_a, zuo_1992_b, croft_2004, meiksin_2004, mcdonald_2005, gontcho_2014, pontzen_2014_a, pontzen_2014_b, daloisio_2018, meiksin_2019, onorbe_2019}.
This is similar to the argument explored by \citet{gaikwad_2023_mfp} in using the probability distribution function of the \lya optical depth to constrain $\lammfp$. 
The probability distribution function of the \lya optical depth does not consider the 2-point clustering, which can be quantified through the auto-correlation function and the power spectrum, which is the Fourier transform of the auto-correlation function, of the \lya forest flux.
Beyond the effect of UVB fluctuations, the power spectrum of the \lya forest flux contrast has been measured at high $z$ and used to constrain the thermal state of the IGM \citep{boera_2019, walther_2019, gaikwad_2021} as well as warm dark matter particle mass \citep{viel_2013, irsic_2017, garzilli_2017}. 

This work is specifically building on \citet{Wolfson_2022} which investigated the effect of a fluctuating UVB on small scales in \lya forest transmission at $z \geq 5.4$.
They found that the \lya forest transmission on small scales will be boosted for small values of $\lammfp$ and that this can be quantified with the \lya forest flux auto-correlation function. 
They used the auto-correlation function to recover $\lammfp$ from simulated mock data. 
The \lya forest flux auto-correlation function has yet to be measured at high-$z$ for observational data. 
Many previous studies have measured the \lya forest flux auto-correlation function at lower redshifts for a wide range of applications \citep{mcdonald_2000, rollinde_2003, becker_2004, dodorico_2006}.

In this paper we use the XQR-30 extended data set to measure the \lya forest flux auto-correlation function. 
We discuss this observational data in Section \ref{sec:data}. 
The details on the data selection and measurement process with a full account of relevant errors are described in Section \ref{sec:methods}. 
We then discuss our resulting measurements in Section \ref{sec:results} and some preliminary comparisons to simulations in Section \ref{sec:sims}.
We summarize our results in Section \ref{sec:conclusion}.

\section{Data} \label{sec:data}

The quasar spectra used in this work are a subset of those presented in \citet{bosman_2021_data}. 
The data reduction was performed and discussed in detail there but will be summarized again in this work for the sake of completeness. 
Additionally, more information on the continuum reconstructions can be found in \citet{bosman_2021_pca}.

\begin{table*}
    \centering
    \caption{
        The extended XQR-30 quasars included in this work. 
        Those with a * represent the extended data set quasars which did not get new spectra in the XQR-30 program. 
        References correspond to: discovery, redshift determination. 
    }
    \label{table:xqr30_qsos}
    \begin{tabular}{lccr}
        Quasar ID         & $z_{\text{qso}}$ & SNR pix$^{-1}$ & Refs.   \\ \hline
        PSO J323+12       & 6.5872           & 35.9           & \citet{mazzucchelli_2017}, \citet{venemans_2020}               \\
        PSO J231-20       & 6.5869           & 42.3           & \citet{mazzucchelli_2017}, \citet{venemans_2020}               \\
        VDES J0224-4711   & 6.5223           & 24.4           & \citet{reed_2017}, \citet{wang_2021}                           \\
        PSO J036+03*      & 6.5405           & 61.4           & \citet{venemans_2015}, \citet{venemans_2020}                   \\
        PSO J1212+0505    & 6.4386           & 55.8           & \citet{mazzucchelli_2017}, \citet{decarli_2018}                \\
        DELS J1535+1943   & 6.3932           & 22.6           & \citet{wang_2019}, \citet{bosman_2021_data}                    \\
        ATLAS J2211-3206  & 6.3394           & 37.5           & \citet{chehade_2018}/\citet{farina_2019}, \citet{decarli_2018} \\
        SDSS J0100+2802*  & 6.3269           & 560.5          & \citet{wu_2015}, \citet{venemans_2020}                         \\
        ATLAS J025-33*    & 6.318            & 127.3          & \citet{carnall_2015}, \citet{becker_2019}                      \\
        SDSS J1030+0524*  & 6.309            & 69.6           & \citet{fan_2001}, \citet{jiang_2007}                           \\
        PSO J060+24       & 6.192            & 49.7           & \citet{banados_2016}, \citet{bosman_2021_data}                 \\
        PSO J065-26       & 6.1871           & 77.9           & \citet{banados_2016}, \citet{venemans_2020}                    \\
        PSO J359-06       & 6.1722           & 68.8           & \citet{wang_2016}, \citet{eilers_2021}                         \\
        % PSO J217-07       & 6.1663           & 33.3           & \citet{banados_2016}, \citet{banados_2016}                     \\
        PSO J217-16       & 6.1498           & 73.0           & \citet{banados_2016}, \citet{decarli_2018}                     \\
        ULAS J1319+0950*  & 6.1347           & 81.7           & \citet{mortlock_2009}, \citet{venemans_2020}                   \\
        CFHQS J1509-1749* & 6.1225           & 43.0           & \citet{willott_2007}, \citet{decarli_2018}                     \\
        PSO J239-07       & 6.1102           & 56.3           & \citet{banados_2016}, \citet{eilers_2021}                      \\
        SDSS J0842+1218   & 6.0754           & 83.2           & \citet{derosa_2011}/\citet{jiang_2015}, \citet{venemans_2020}  \\
        ATLAS J158-14     & 6.0685           & 60.3           & \citet{chehade_2018}, \citet{eilers_2021}                      \\
        VDES J0408-5632   & 6.0345           & 86.6           & \citet{reed_2017}, \citet{reed_2017}                           \\
        SDSS J1306+0356*  & 6.033            & 65.3           & \citet{fan_2001}, \citet{venemans_2020}                        \\
        ATLAS J029-36     & 6.021            & 57.1           & \citet{carnall_2015}, \citet{becker_2019}                      \\
        SDSS J2310+1855   & 6.0031           & 113.4          & \citet{jiang_2016}, \citet{wang_2013}                          \\
        PSO J007+04       & 6.0015           & 54.4           & \citet{jiang_2015}/\citet{banados_2014}, \citet{venemans_2020} \\
        ULAS J0148+0600*  & 5.998            & 152.0          & \citet{jiang_2015}, \citet{becker_2019}                        \\
        SDSS J0818+1722*  & 5.997            & 132.1          & \citet{fan_2006}, \citet{becker_2019}                          \\
        PSO J029-29       & 5.984            & 65.6           & \citet{banados_2016}, \citet{banados_2016}                     \\
        PSO J108+08       & 5.9485           & 104.8          & \citet{banados_2016}, \citet{banados_2016}                     \\
        PSO J183-12       & 5.917            & 61.8           & \citet{banados_2014}, \citet{bosman_2021_data}                 \\
        PSO J025-11       & 5.844            & 50.6           & \citet{banados_2016}, \citet{bosman_2021_data}                 \\
        PSO J242-12       & 5.837            & 22.9           & \citet{banados_2016}, \citet{bosman_2021_data}                 \\
        PSO J065+01       & 5.833            & 25.1           & \citet{dodorico_2023_xqr30}, \citet{bosman_2021_data}          \\
        SDSS J0836+0054*  & 5.804            & 73.8           & \citet{fan_2001}, \citet{bosman_2021_data}                     \\
        PSO J308-27       & 5.7985           & 53.2           & \citet{banados_2016}, \citet{dodorico_2023_xqr30}              \\
        SDSS J0927+2001*  & 5.7722           & 53.8           & \citet{fan_2006}, \citet{wang_2010}     
    \end{tabular}
\end{table*}

All of the observations used in this work comes from the XQR-30 program\footnote{\url{https://xqr30.inaf.it/}} \citep[1103.A0817(A), ][]{dodorico_2023_xqr30}, which consists of a sample of 30 very luminous quasars at $z \gtrsim 5.8$ observed with the X-Shooter instrument \citep{vernet_2011} on the Very Large Telescope. 
We use 24 quasars from the XQR-30 sample which do not show strong broad absorption lines (BALs) that would create issues in the modelling of the intrinsic continuum \citep{Bischetti_2022} and could also possibly contaminate the \lya forest region. 
Three additional spectra (PSO J231-20, ATLAS J2211-3206, and SDSS J2310+1855) were identified as hi-BALs so we exclude regions of the spectra where there is possible strong OVI contamination ($7770\text{\AA} < \lamobs < 7870\text{\AA}$, $\lamobs < 7280\text{\AA}$, and $\lamobs < 6700\text{\AA}$ respectively). 
All XQR-30 spectra have signal-to-noise ratios (SNRs) larger than 20 per $\SI{10}{\kilo\meter\per\second}$ pixel measured over $1165\text{\AA} < \lamrest < 1170\text{\AA}$ (Table \ref{table:xqr30_qsos}). 
In addition to the 24 XQR-30 quasars, we use 11 archival X-Shooter spectra that are from the extended XQR-30 sample \citep{dodorico_2023_xqr30}.
These spectra have SNR > 40 per $\SI{10}{\kilo\meter\per\second}$ pixel from the literature (Table \ref{table:xqr30_qsos}, marked with *).
The extended XQR-30 sample has a median effective resolving power over all 42 quasars of $R \simeq 11400$ and 9800 in the visible ($5500\text{\AA} < \lamobs < 10200\text{\AA}$) and infrared arm ($10200\text{\AA} < \lamobs < 24800\text{\AA}$), respectively \citet{dodorico_2023_xqr30}.

All quasars are reduced with the same procedure. 
Observations are first flat-fielded and sky-subtracted following the method of \citet{kelson_2003}. 
The spectra are extracted \citep{horne_1986} separately for the visible and infrared arms of the instrument which are then stitched together over the $10110\text{\AA} < \lamobs < 10130\text{\AA}$ spectral window.
The infrared spectrum is re-scaled to match the observed mean flux in the optical arm. 
The spectrum is then interpolated over the overlap window in order to minimize the risk of creating an artificial step in the spectrum between the arms to which the continuum-fitting method may be non-linearly sensitive \citep[see discussion in][]{bosman_2021_data}. 
The reduction routines are described in more detail in \citet{becker_2009}.
Further details are presented in \citet{dodorico_2023_xqr30}.

\begin{figure*}
    \includegraphics[width=2\columnwidth]{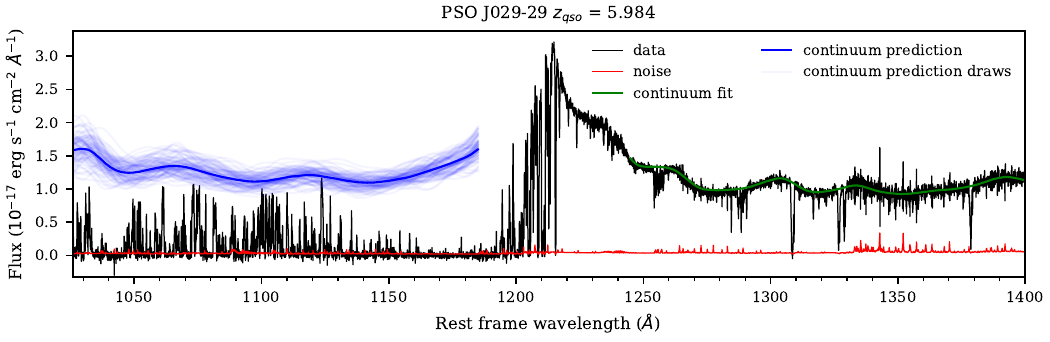}
    \caption{
        The X-Shooter spectrum of the \lya transmission region for the quasar PSO J029-29 from the XQR-30 sample. 
        The noise vector is shown in red and the PCA-reconstructed continuum is shown in blue. 
        The light blue lines show draws of the continuum reconstruction with the appropriate scatter from the covariance matrix of the PCA reconstruction.
        The pixel scale is \SI{10}{\kilo\meter\per\second} and the SNR of the \lya region (reconstruction divided by uncertainty) is SNR = 50.6. 
    }
    \label{fig:one_spectra}
\end{figure*}

An example spectrum from the program is shown in Figure \ref{fig:one_spectra} for PSO J029-29. 
The black shows the reduced XQR-30 spectra and the red shows the noise vector. 
The intrinsic continuum reconstructed with the method described in Section \ref{sec:continuum fitting} is shown by the solid blue line, while the continuum fit to the red side of the quasars emission is shown in green. 
The light blue lines show draws of the continuum reconstruction with the appropriate scatter from the covariance matrix of the PCA reconstruction.
The sampling procedure for these draws are also discussed in Section \ref{sec:continuum fitting}.

\section{Methods} \label{sec:methods}

\subsection{Continuum reconstruction} \label{sec:continuum fitting}

For each quasar, the continuum, $F_{\text{cont}}(\lamrest)$, was reconstructed using Principal Component Analysis (PCA).  
To do this we consider both the red side ($\lamrest > 1280\text{\AA}$) and the blue side ($\lamrest < 1220\text{\AA}$) of the quasar continuum with respect to the \lya emission.
At low-$z$, both sides of the quasar continuum are transmitted through the IGM, as the IGM is mainly ionized. 
Thus we can use PCA to find the optimal linear decomposition of both the red side and the blue side of the low-$z$ quasar continuum, then construct an optimal mapping between the the linear coefficients from the two decompositions. 
At high-$z$, the red side of quasar continua will be transmitted while the blue side is absorbed by remaining neutral hydrogen in the IGM, see e.g. Figure \ref{fig:one_spectra}. 
We can thus get the PCA decomposition for the red side of the continuum then use the optimal mapping, determined from low-$z$ quasars, to predict the blue side coefficients and thus the 
continuum \citep{francis_1992, yip_2004}. 
This method has been historically used to get the continuum for the \lya forest in \citet{suzuki_2005} then was further expanded, for example by: \citet{mcdonald_2005, paris_2011, davies_2018_b, davies_2018_log_pca, durovcikova_2020}. 
Previously, \citet{bosman_2021_pca} determined the most accurate PCA method and \citet{bosman_2021_data} further improved this method with the log-PCA approach of \citet{davies_2018_b, davies_2018_log_pca}. 

This work uses the same reconstructions that were generated for \citet{bosman_2021_data} using the log-PCA approach. 
The PCA consists of 15 red-side components and 10 blue-side components. 
The training set was 4597 quasars from the SDSS-III Baryon Oscillation Spectroscopic Survey \citep[BOSS,][]{dawson_2013} and the SDSS-IV Extended BOSS \citep[eBOSS,][]{dawson_2016} at $2.7 < z < 3.5$ with SNR > 7. 
Intrinsic continua were obtained automatically using a modified version of the method of \citet{dallaglio_2008}, originally based on the procedures outlined in \citet{young_1979} and \citet{carswell_1982}.
These continua are re-normalized so that they match the observed mean \lya transmission at $z \sim  3$ that was measured from high-resolution spectra \citep{faucher-giguere_2008, becker_2013} to prevent bias from the low spectral resolution of the SDSS spectrograph \citep[as described in][]{dallaglio_2009}.
The reconstructions were tested with an independent set of 4597 quasars from eBOSS. 
As described in \citet{bosman_2021_data}, this testing revealed that there is no bias in reconstructing the blue-side emission lines and that the method predicts the underlying continuum within 8\%.
The reconstruction error on this testing set gives us the mean, $\boldsymbol{\mu_{\text{cont}}}$, and covariance, $\boldsymbol{\Sigma_{\text{cont}}}$, of the PCA reconstruction as shown in Figure 2 of \citet{bosman_2021_data}. 

Going forward, we always forward-model the full wavelength-dependent uncertainties from the reconstruction of $F_{\text{cont}}(\lamrest)$ into all measurements and model comparisons. 
We do this by randomly drawing realizations of the continuum error, $\boldsymbol{E_{\text{cont}}} \sim N(\boldsymbol{\mu_{\text{cont}}}, \boldsymbol{\Sigma_{\text{cont}}})$, where $N$ is the normal distribution. 
We create a realization of the predicted continuum with this error, $\boldsymbol{C_{\text{pred}}}$, from the fit quasar continuum, $\boldsymbol{C_{\text{fit}}}$, via: 
\begin{equation}
    \boldsymbol{C_{\text{pred}}} = \boldsymbol{C_{\text{fit}}} \times \boldsymbol{E_{\text{cont}}}.
\end{equation}
We use 500 of these continuum draws to analyze each quasar's spectrum.
When we performed bootstrap re-sampling as described in Section \ref{sec:bootstrap}, each draw uses a random selection of these 500 continua. 
Figures showing all PCA fits and blue-side predictions are shown in \citet{zhu_2021}.
% and the PCA fits for XQR-30 spectra have been made public with the first XQR-30 data release \citep{dodorico_2023_xqr30}.

\subsection{Pixel masking} \label{sec:pixel_mask}

We want to use flux from the quasar continuum that exclusively corresponds to \lya forest absorption. 
To do this, we only use wavelengths larger than the \lyb emission at the redshift of the quasar, or $\lamrest > 1026\text{\AA}$.
Additionally, we want to exclude the quasars proximity zone, which is the region close to the quasar where the IGM has been ionized by the quasar's own emission and the transmission is enhanced.  
For this reason, we consider $\lamrest < 1185\text{\AA}$ following \citet{bosman_2021_data} which corresponds to $\sim \SI{7650}{\kilo\meter\per\second}$ from emission at $z \sim 6$.
This is a conservative estimate based on \citet{bosman_2018}, which found no effect on the \lya transmission in spectral stacks over this wavelength.

The data reduction procedure should automatically reject outlier pixels.
However, we check for and exclude anomalous pixels that meet either of the following conditions: the SNR at the unabsorbed continuum level is < 2 per pixel or if pixels have negative flux at $> 3\sigma$ significance. 
This excludes 0\% of pixels for the SNR cut at all redshifts and $0.07 - 0.47\%$ of pixels for the negative flux cut depending on redshift.

\subsection{DLA exclusion} \label{sec:dla_mask}

Damped \lya absorption systems (DLAs) are intervening systems along quasar sightlines with hydrogen column densities $N_{\text{HI}} \geq 10^{20.3}$ cm$^{-2}$.
These systems result in significant damping wings in the \lya absorption profile \citep{wolfe_2005, rafelski_2012}. 
DLAs in quasar spectra at $z \gtrsim  6$ can cause complete absorption of \lya transmission over $\Delta v = \SI{2000}{\kilo\meter\per\second}$ and additional suppression over $\Delta v \gtrsim \SI{5000}{\kilo\meter\per\second}$ intervals \citep{dororico_2018, banados_2019, davies_2020}.
DLAs can arise in the circumgalactic medium (CGM) of galaxies which are not typically included in reionization simulations, including those discussed in Section \ref{sec:sims}.
For this reason, we attempt to remove DLAs from our observations based on the presence of metals in the spectra.
This does leave open the possibility that DLAs from neutral patches of the IGM remain in our observations. 

We remove DLAs by identifying and masking out their locations in our spectra. 
The detection of $z \gtrsim 5$ DLAs relies on the identification of associated low-ionization metal absorption lines, since the \lya absorption from the DLA may not be able to be distinguished from the highly-opaque IGM.
The typical transitions are CII, OI, SiII, and MgII.  
DLA metallicities at $z \gtrsim 5$ vary so even relatively weak metal absorption could indicate a DLA. 
The identification of intervening metal absorbers in the extended XQR-30 sample has been described in detail in \citet{davies_r_2023}.
Due to the high SNR of the X-Shooter spectra, we expect to be > 90\% complete to absorption corresponding to $\log N_{\text{MgII}}/$cm$^{-2} \gtrsim 13$. 

We adopt the following criteria for our masks, following \citet{bosman_2021_data}. 
We mask the central $\Delta v = \SI{3000}{\kilo\meter\per\second}$ for systems with metal column densities $\log N_{\text{CII}}$/cm$^{-2} > 13$, $\log N_{\text{OI}}$/cm$^{-2} > 13$, or $\log N_{\text{SiII}}$/cm$^{-2} > 12.5$, measured through the $\lamrest = 1334.53$\AA, $1302.16$\AA, and $1526$\AA \ transitions, respectively. 
When none of these ions are accessible, we also exclude the central $\Delta v = \SI{3000}{\kilo\meter\per\second}$ for systems with $\log N_{\text{MgII}}$/cm$^{-2} > 13$ based on the high rates of co-occurrence of the MgII 2796.35, 2803.53\AA \ doublet \citep{cooper_2019}. 
We exclude a larger window of $\Delta v = \SI{5000}{\kilo\meter\per\second}$ around intervening systems with $\log N_{\text{OI, CII, SiII, MgII}}$/cm$^{-2} > 14$ due to the likely presence of extended damping wings.
We do not exclude systems based on the presence of highly ionized ions alone (e.g. C IV, Si IV) since the corresponding gas is likely highly ionized \citep{cooper_2019}. 

We investigate the effect of this mask on the measurement of the auto-correlation function in Appendix \ref{appendix:dla}.

\subsection{Resulting normalized flux} \label{sec:norm flux and masks}

\begin{figure*}
	\includegraphics[width=2\columnwidth]{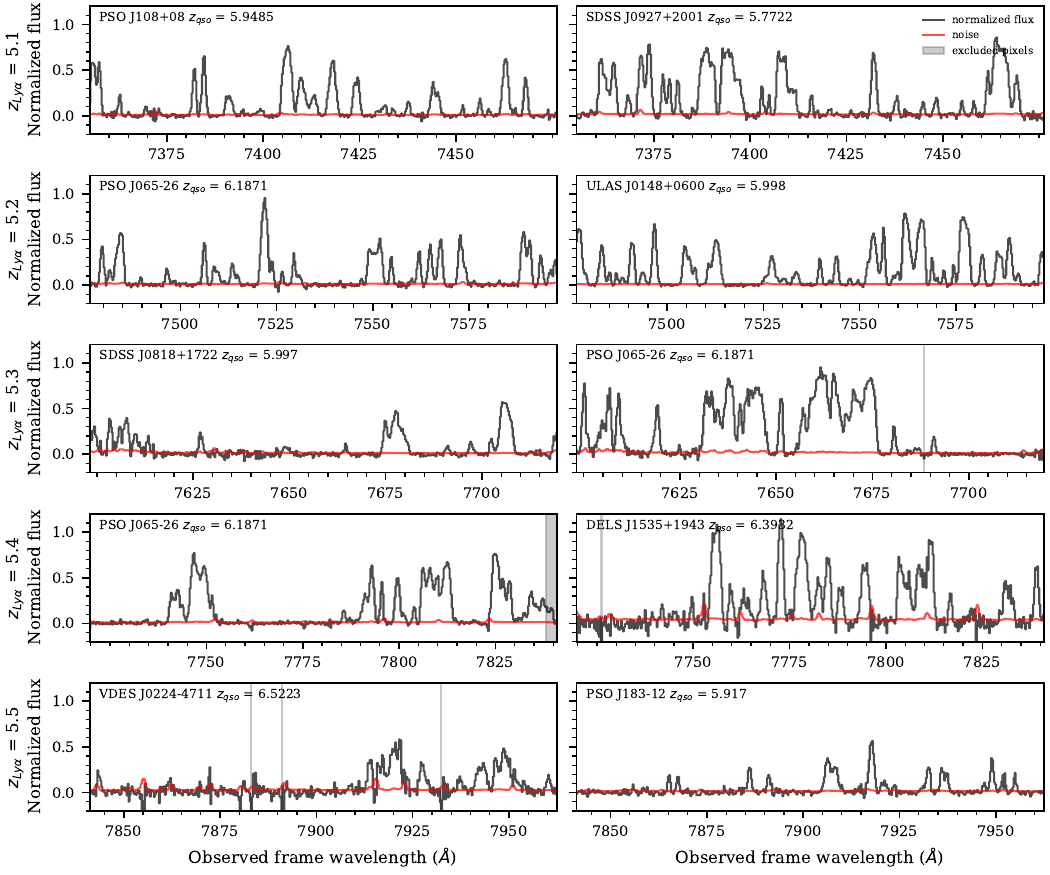}
    \caption{
        This figure shows the continuum normalized flux for two randomly selected quasars at five values of $z$ of the \lya forest from $5.1 \leq z \leq 5.5$.
        These sections are centered on the given $z$ and span $\Delta z = 0.05$. 
        The continuum normalized flux is shown in black with the continuum normalized uncertainty in red.
        The shaded regions indicate excluded pixels based on the masking procedure described in Section \ref{sec:pixel_mask} and \ref{sec:dla_mask}.  
        Each row shares the same y-axis to demonstrate the decrease in $\langle F \rangle$ with increasing $z$ (down the rows). 
        Note that the normalized flux for all the quasars considered in each measurement can be found in the online supplementary material. 
    }
    \label{fig:norm_flux_low}
\end{figure*}

\begin{figure*}
	\includegraphics[width=2\columnwidth]{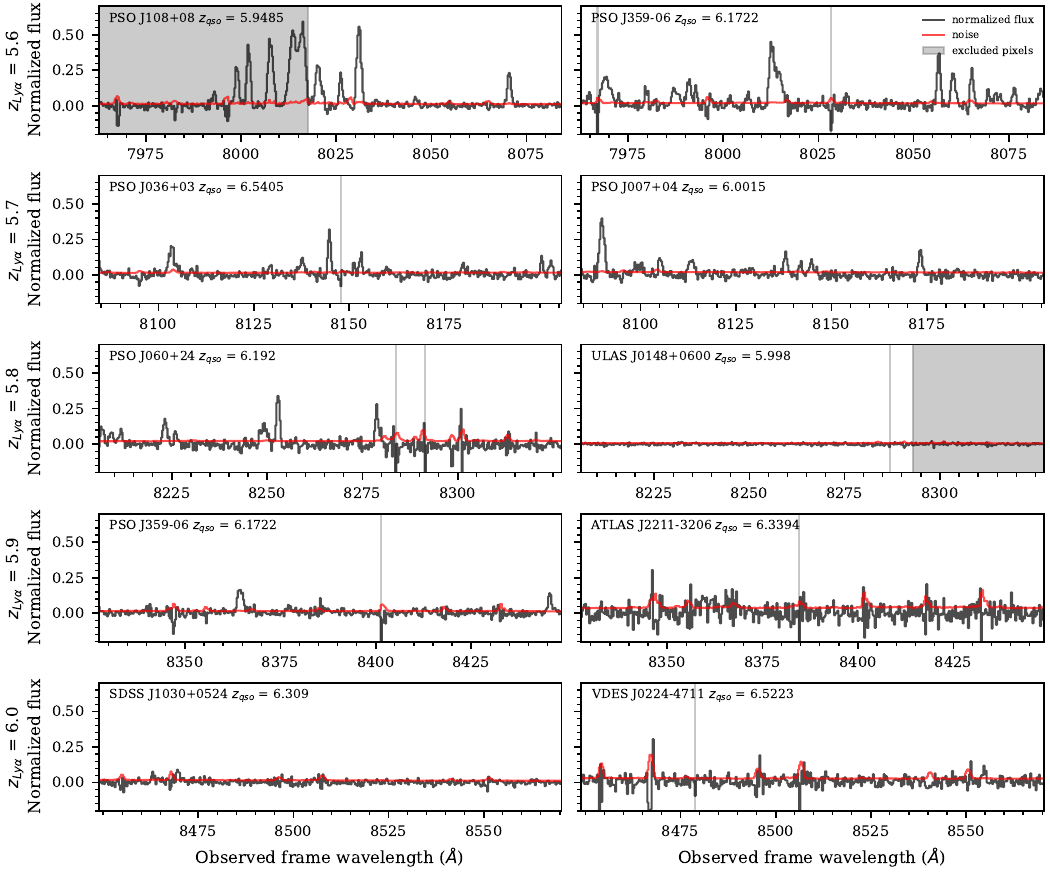}
    \caption{
        The same as Figure \ref{fig:norm_flux_low} except for $5.6 \leq z \leq 6.0$. 
        The y-axis spans a smaller range than that in Figure \ref{fig:norm_flux_low}. 
    }
    \label{fig:norm_flux_high}
\end{figure*}

After combining the masks of the bad pixels discussed in Section \ref{sec:pixel_mask} and the DLAs discussed in Section \ref{sec:dla_mask} we only considered sightlines that maintain at least $10\%$ of the pixels in each redshift bin.
Only using spectra that maintain at least 10\% of pixels limits noisy contributions to the measurement from short spectra that may only consist of one transmission spike. 
Two random examples of the normalized flux from quasars in our sample at each redshift are shown in Figures \ref{fig:norm_flux_low} and \ref{fig:norm_flux_high}. 
The normalized flux for all the sightlines used in each redshift bin can be found in the online supplementary material, which demonstrate the variance between the sightlines at a given redshift. 

Figure \ref{fig:norm_flux_low} shows the normalized flux for $5.1 \leq z \leq 5.5$.
Each row has the same $z$ and each column shows a random quasar sightline. 
The value of $z$ increases down the rows. 
These plots show the normalized flux in black and the 1$\sigma$ noise for the normalized flux in red.
The grey shaded regions represent the excluded regions due to the mask from the bad pixels discussed in Section \ref{sec:pixel_mask} and the DLAs discussed in Section \ref{sec:dla_mask}. 
The y-axis is fixed to the same range in each panel of the figure, which allows the figure to demonstrate a rough trend of decreasing $\langle F \rangle$ with increasing $z$. 

Figure \ref{fig:norm_flux_high} shows the same normalized flux plot, this time for $5.6 \leq z \leq 6.0$. 
Again the y-axis in each row is fixed at the same range such to demonstrate the decrease in $\langle F \rangle$ with increasing $z$.
The y-axis limits are not the same between Figures \ref{fig:norm_flux_low} and \ref{fig:norm_flux_high}. 
Both of the random sightlines shown at $z = 6$ have very limited transmission, which highlights the difficulty in making statistical measurements of the \lya forest at high redshifts.

\section{Results} \label{sec:results}

\subsection{Mean flux} \label{sec:mean_flux}

\begin{table}
    \centering
    \begin{tabular}{ccc}
        $z$ & $N_{\text{los}}$ & $\langle F \rangle$ \\ \hline
        5.1 & 24               & $0.1456 \pm 0.0075$ \\
        5.2 & 29               & $0.1314 \pm 0.0072$ \\
        5.3 & 29               & $0.1097 \pm 0.0087$ \\
        5.4 & 33               & $0.0830 \pm 0.0086$ \\
        5.5 & 34               & $0.0567 \pm 0.0055$ \\
        5.6 & 34               & $0.0474 \pm 0.0053$ \\
        5.7 & 29               & $0.0269 \pm 0.0044$ \\
        5.8 & 26               & $0.0181 \pm 0.0035$ \\
        5.9 & 15               & $0.0089 \pm 0.0018$ \\
        6.0 & 14               & $0.0090 \pm 0.0023$
    \end{tabular}
    \caption{
        The second column lists the numbers of lines of sight at each $z$ in our sample. 
        The third column reports the mean flux, $\langle F \rangle$, value that was directly computed from this sample. 
        The error on $\langle F \rangle$ comes from bootstrap re-sampling of the sightlines.
    }
    \label{table:quasar_los_flux}
\end{table}

\begin{figure}
	\includegraphics[width=\columnwidth]{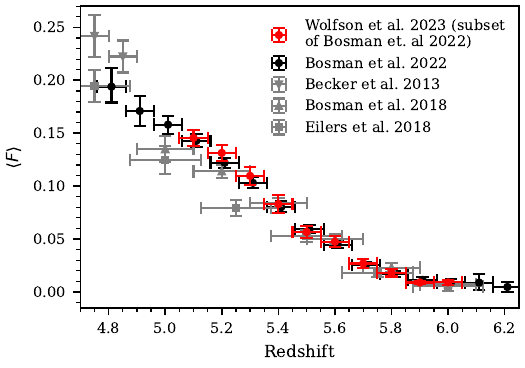}
    \caption{
        Recent measurements of the average \lya transmission, $\langle F \rangle$, at high-$z$.
        The measured $\langle F \rangle$ from this work are shown in red. 
        This is computed directly by taking the average of the non-excluded normalized flux values from the masks created as discussed in Sections \ref{sec:pixel_mask} and \ref{sec:dla_mask}.
        The errors come from bootstrap re-sampling the quasar sightlines. 
        Note that the measurement shown in red comes from a subset of the quasar sightlines used in \citet{bosman_2021_data} which are plotted in black.
        Additional data points from previous works are shown in grey over the same $z$ range \citep{becker_2013, bosman_2018, eilers_2018}.
    }
    \label{fig:measured_flux}
\end{figure}

The mean flux in this paper was calculated as the average of the normalized flux values for the non-excluded pixels as shown in Figures \ref{fig:norm_flux_low} and \ref{fig:norm_flux_high}. 
The resulting values are reported in Table \ref{table:quasar_los_flux} and plotted as a function of redshift in Figure \ref{fig:measured_flux}. 
The error on the $\langle F \rangle$ values were computed by bootstrap re-sampling the quasar sightlines considered at each $z$ for 500,000 data set realizations and computing the variance on these values. 
See Section \ref{sec:bootstrap} for more information on how the bootstrap realizations were generated.

Figure \ref{fig:measured_flux} shows the $\langle F \rangle$ values computed in this work in red, the previous measurement of \citet{bosman_2021_data} in black, and the measurements of \citet{bosman_2018, becker_2013, eilers_2018} in grey. 
Our measurement is in agreement with that from \citet{bosman_2021_data}, as is expected since the data used here is a subset of that used in that work and our method is the same.
In addition, we use the same continuum reconstruction and masking procedure as in \citet{bosman_2021_data}.
At $z = 5.1$ and $z = 5.2$ our measurement appears greater than that from \citet{bosman_2021_data}, but the data set we considered is much smaller and the measurements are consistent within the error bars. 
A discussion of the agreement of $\langle F \rangle$ with previous work can be found in \citet{bosman_2021_data}.

\subsection{Auto-correlation Function}

The auto-correlation function of the flux ($\xi_F (\Delta v)$) is defined as
\begin{equation}
    \xi_F (\Delta v) = \langle F(v) F(v + \Delta v) \rangle
    \label{eq:autocorr}
\end{equation}
where $F(v)$ is the flux of the \lya forest and the average is performed over all pairs of pixels at the same velocity lag ($\Delta v$). 
The pixels that have been masked as discussion in Sections \ref{sec:pixel_mask} and \ref{sec:dla_mask} are not used when computing the auto-correlation function for each quasar. 
See Appendix \ref{appendix:dla} for a discussion of the effect of the DLA exclusion on the measurement of the auto-correlation function. 
Note that different quasar sightlines will have a different number of pixel pairs contributing to the same velocity bin.
To account for this we use a weighted average when combining the different quasar sightlines, where the numbers of pixel pairs per bin are the weights. 
The number count of pixel pairs contributing to each auto-correlation function bin is output during the auto-correlation function computation.

We compute the auto-correlation function with the following consideration for the velocity bins. 
We start with the left edge of the smallest bin to be $\SI{40}{\kilo\meter\per\second}$ and use linear bins with a width of $\SI{40}{\kilo\meter\per\second}$ up to $\SI{280}{\kilo\meter\per\second}$. 
Then we switch to logarithmic bin widths where $\log_{10}(\Delta v) = 0.058$ out to a maximal distance of $\SI{2700}{\kilo\meter\per\second}$. 
This results in 22 velocity bins considered where the first 6 have linear spacing. 
The center of our smallest bin was $\SI{60}{\kilo\meter\per\second}$ and our largest bin was $\SI{2223}{\kilo\meter\per\second}$, which corresponds to $\sim 16$ cMpc h$^{-1}$ at $z = 5.5$. 
We chose to use linear bins on the smallest scales because the effect of $\lammfp$ is greatest on small scales and these scales already have access to the most pixel pairs which reduces noise. 
Larger scales are more sensitive to $\langle F \rangle$ than $\lammfp$ so having fewer bins here is not as important.
In addition, there are fewer pixel pairs at large scales to begin with so using larger bins will increase the pixel pairs per bin and reduce noise. 

% Conventionally, the flux contrast field, $(F - \langle F \rangle) / \langle F \rangle$, is used when measuring statistics of the \lya forest. 
% Here, we use the flux since $\langle F \rangle$ is small and has large uncertainties at high-$z$, which can be seen in the values of $\langle F \rangle$ reported in Table \ref{table:quasar_los_flux}.
% This way we avoid dividing by a small number, which could potentially blow up the values of the auto-correlation function. 

\begin{figure*}
	\includegraphics[width=2\columnwidth]{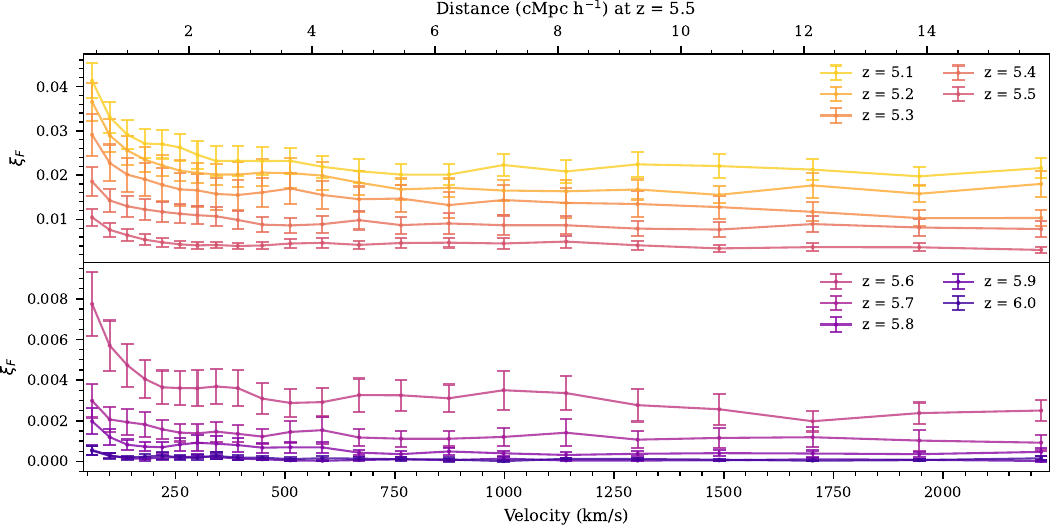}
    \caption{
        The auto-correlation function of \lya transmission in ten redshift bins for XQR-30 data. 
        The top panel shows the lower $z$ bins, $5.1 \leq z \leq 5.5$ while the lower panel shows the higher $z$ bins, $5.6 \leq z \leq 6.0$. 
        The main trend seen in these plots is the evolution of $\langle F \rangle$ which is very small at high-$z$.
    }
    \label{fig:measured_autocorr}
\end{figure*}

\begin{figure*}
	\includegraphics[width=2\columnwidth]{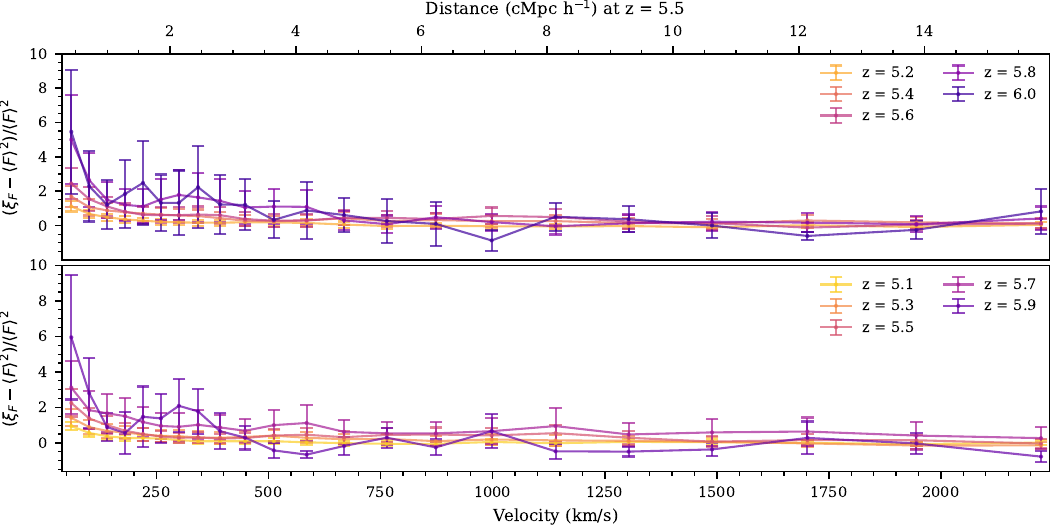}
    \caption{
        The auto-correlation function of \lya transmission normalized and shifted by the mean transmission, $\langle F \rangle$, in ten redshift bins for XQR-30 data. 
        This is equivalent to the auto-correlation function of the flux density field. 
        The errors are computed by propagating the statistical uncertainty from bootstrap re-sampling both the auto-correlation function and $\langle F \rangle$. 
        This is split into two panels for visual clarity, so as to not overcrowd the panels. 
        The top panel has $z = 5.1, 5.3, 5.5, 5.7, 5.9$ while the bottom panel has $5.2, 5.4, 5.6, 5.8, 6.0$. 
        This figure makes the trend of higher redshift bins having larger boosts of the auto-correlation function on small scales when dividing out the flux evolution more visible. 
    }
    \label{fig:measured_autocorr_norm}
\end{figure*}

Previously, \citet{Wolfson_2022} demonstrated the sensitivity of the auto-correlation function to $\lammfp$ for mock data at $z \geq 5.4$. 
Generally, they found that shorter $\lammfp$ values cause a greater boost in the auto-correlation function on the smallest scales. 
We compute the auto-correlation functions of the XQR-30 data set discussed in Section \ref{sec:data}. 
The measured auto-correlation function from the extended XQR-30 data set can be seen in Figures \ref{fig:measured_autocorr} and \ref{fig:measured_autocorr_norm}. 
The errors on these plots come from bootstrap sampling of the quasar sightlines when computing the mean auto-correlation function and will be discussed in more detail in Section \ref{sec:bootstrap}. 
The first few velocity bins of the final measurement with error from the diagonal of the covariance matrix estimated via bootstrap re-sampling are in Table \ref{table:autocorr_incomplete}. 
The full measurement, error bars, as well as the full bootstrap covariance matrices for each redshift are available to download online\footnote{\url{https://github.com/mollywolfson/lya\_autocorr/}}.

\begin{table*}
    \centering
    \caption{
        The table lists the auto-correlation function measurement for the first six bins of the auto-correlation function at all $z$ with errors from the diagonal of the covariance matrix estimated from bootstrap re-sampling.
        The full measurement values of the auto-correlation function at all $z$ can be found online.
    }
    \begin{tabular}{|c|ccccccc|}
        \hline
        \multicolumn{1}{|l|}{$\boldsymbol{z}$} & \multicolumn{7}{c|}{\textbf{Central velocity (km s$^{-1}$)}}                                                                                                    \\ \hline
                                           & 60                     & 100                    & 140                      & 180                      & 220                    & 260                      & ... \\ \cline{2-8} 
        5.1                                & $0.0413 \pm 0.0040$    & $0.0331 \pm  0.0035$   & $0.0291 \pm  0.0033$     & $0.0271 \pm  0.0032$     & $0.0270 \pm  0.0032$   & $0.0262 \pm  0.0031$     & ... \\
        5.2                                & $0.0365 \pm  0.0043$   & $0.0290 \pm  0.0037$   & $0.0256 \pm  0.0032$     & $0.0235 \pm  0.0028$     & $0.0222 \pm  0.0025$   & $0.0210 \pm  0.0024$     & ... \\
        5.3                                & $0.0291 \pm 0.0047$    & $0.0227 \pm  0.0042$   & $0.0201 \pm  0.0038$     & $0.0191 \pm  0.0037$     & $0.0178 \pm  0.0037$   & $0.0168 \pm  0.0036$     & ... \\
        5.4                                & $0.0185 \pm 0.0032$    & $0.0143 \pm  0.0026$   & $0.0130 \pm  0.0024$     & $0.0122 \pm  0.0023$     & $0.0117 \pm  0.0023$   & $0.0113 \pm  0.0022$     & ... \\
        5.5                                & $0.0105 \pm 0.0020$    & $0.0076 \pm  0.0016$   & $0.0064 \pm  0.0013$     & $0.0054 \pm  0.0012$     & $0.0048 \pm  0.0010$   & $0.00436 \pm  0.00081$   & ... \\
        5.6                                & $0.0078 \pm 0.0016$    & $0.0057 \pm  0.0012$   & $0.0047 \pm  0.0011$     & $0.00406 \pm  0.00093$   & $0.00365 \pm  0.00083$ & $0.00361 \pm  0.00084$   & ... \\
        5.7                                & $0.00298 \pm 0.00082$  & $0.00206 \pm  0.00062$ & $0.00193 \pm  0.00065$   & $0.00182 \pm  0.00060$   & $0.00158 \pm  0.00048$ & $0.00141 \pm  0.00044$   & ... \\
        5.8                                & $0.00197 \pm  0.00065$ & $0.00120 \pm  0.00040$ & $0.00082 \pm  0.00026$   & $0.00072 \pm  0.00022$   & $0.00070 \pm  0.00027$ & $0.00083 \pm  0.00032$   & ... \\
        5.9                                & $0.00055 \pm  0.00023$ & $0.00030 \pm  0.00013$ & $0.000150 \pm  0.000045$ & $0.000124 \pm  0.000085$ & $0.00020 \pm  0.00012$ & $0.000189 \pm  0.000095$ & ... \\
        6.0                                & $0.00053 \pm  0.00023$ & $0.00027 \pm  0.00014$ & $0.000180 \pm  0.000096$ & $0.00023 \pm  0.00014$   & $0.00028 \pm  0.00017$ & $0.00019 \pm  0.00012$   & ... \\ \hline
    \end{tabular}
    \label{table:autocorr_incomplete}
\end{table*}

Figure \ref{fig:measured_autocorr} has two panels that show the auto-correlation function of this data set at different $z$. 
The top panel shows $5.1 \leq z \leq 5.5$ while the bottom panel shows $5.6 \leq z \leq 6.0$. 
They are shown in two different panels in order to better accommodate the dynamic range of the auto-correlation function over our range of $z$.
The overall amplitude of the auto-correlation function of the flux is set by $\langle F \rangle^2$, which decreases with increasing $z$.

In order to better visually demonstrate the differences in the shape of the auto-correlation function on small scales, we also plot the measured auto-correlation function normalized and shifted by $\langle F \rangle^2$ at each $z$ in Figure \ref{fig:measured_autocorr_norm}.
Note that the $\langle F \rangle$ value used is redshift dependent and is reported in Table \ref{table:quasar_los_flux}.
This is equivalent to the auto-correlation function of the flux density field. 
The color of the normalized auto-correlation function at each $z$ matches those from Figure \ref{fig:measured_autocorr}. 
This has been split into two panels for visual clarity to more easily see the behavior in each redshift bin. 
The top panel has $z = 5.1, 5.3, 5.5, 5.7, 5.9$ while the bottom panel has $5.2, 5.4, 5.6, 5.8, 6.0$. 
By looking at the smallest scales, $v < \SI{500}{\kilo\meter\per\second}$ or $x < 4$ cMpc h$^{-1}$ at $z = 5.5$, there is a trend of increasing small-scale values of the auto-correlation function with increasing redshift.
For example, the lines for $5.8 \leq z \leq 6.0$ have the greatest auto-correlation value (in shades of purple).
Note that these points have the largest error bars, likely caused by both the limited number of sightlines and the low transmission at these redshifts. 
Both $\langle F \rangle$ and $\lammfp$ affect the small scale boost in the auto-correlation function. 
Smaller $\langle F \rangle$ will lead to larger fluctuations in the flux contrast field and thus a boost on the small scales. 
\citet{Wolfson_2022} found that shorter $\lammfp$ values also cause a boost in the auto-correlation function on the smallest scales. 
These effects are not completely degenerate since the overall auto-correlation function shape differs as shown in the forecast measurements of \citet{Wolfson_2022}.  

\begin{figure}
	\includegraphics[width=\columnwidth]{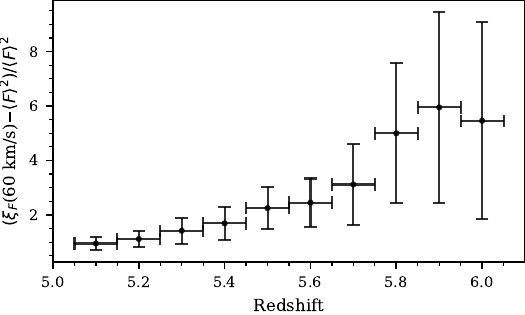}
    \caption{
        The value of the first bin of the auto-correlation function of \lya transmission normalized and shifted by the mean transmission, $\langle F \rangle$, as a function of redshift. 
        The errors are computed by propagating the statistical uncertainty from bootstrap re-sampling both the auto-correlation function and $\langle F \rangle$. 
        These values are also shown in Figure \ref{fig:measured_autocorr_norm}.
        There is a general trend of increasing value with redshift, though the errors also increase. 
        The highest redshift values are consistent with no evolution. 
    }
    \label{fig:measured_autocorr_norm_first_bin}
\end{figure}

We isolate the redshift evolution of the smallest velocity bin ($\SI{60}{\kilo\meter\per\second}$) of the normalized auto-correlation function in Figure \ref{fig:measured_autocorr_norm_first_bin}. 
Again, the $\langle F \rangle$ value used is redshift dependent and is reported in Table \ref{table:quasar_los_flux}.
The errors are computed by propagating the statistical uncertainty from bootstrap re-sampling both the auto-correlation function and $\langle F \rangle$. 
In general these values increase with redshift, which is expected from decreasing $\langle F \rangle$ as well as $\lammfp$.
However, the errors also increase with redshift and the values at the highest redshift are consistent with each other within errors.

\subsection{Bootstrap Covariance Matrices} \label{sec:bootstrap}

In order to calculate the error on $\langle F \rangle$ and the auto-correlation functions we used bootstrapping re-sampling. 
To compute the values we performed averages over $N_{\text{boot}} = 500000$ realizations of the data set. 
Each realization is a random selection of $N_{\text{los}}$ quasars with replacement. 
In addition, each choice of quasar goes along with a choice of the 500 continuum realizations that were generated as described at the end of Section \ref{sec:continuum fitting}. 
The computed mean flux for the $i$th sample is thus $\langle F \rangle_{i}$ and the error on $\langle F \rangle$, $\sigma_F$ is:
\begin{equation}
    \sigma_{F} = \left( \frac{1}{N_{\text{boot}}-1} \sum_{i=1}^{N_{\text{boot}}} \left( \langle F \rangle_{i} - \langle F \rangle \right)^2 \right)^{1/2}.
    \label{eq:flux_boot_error}
\end{equation}
These errors are reported in Table \ref{table:quasar_los_flux} and shown in Figure \ref{fig:measured_flux}. 

For the auto-correlation function, $\xi$, we compute the entire bootstrap covariance matrix, not only the diagonal error. 
Again we chose $N_{\text{boot}}$ realizations of the observed data set by randomly selecting $N_{\text{los}}$ quasars with replacement each with their own random selection of the continuum realization.
For any given bootstrap realization we computed the average of the auto-correlation function over the chosen sightlines to construct a realization of the average auto-correlation function, $\xi_i$.
The covariance matrix was then computed by averaging over the ensemble of bootstrap realizations in the following way:
\begin{equation}
    \boldsymbol{\Sigma_{\text{boot}}} = \frac{1}{N_{\text{boot}}-1} \sum_{i=1}^{N_{\text{boot}}}(\boldsymbol{\xi}_i - \boldsymbol{\xi}_{\text{data}})(\boldsymbol{\xi}_i - \boldsymbol{\xi}_{\text{data}})^{\text{T}}.
    \label{eq:boot_covariance}
\end{equation}
For visualization purposes, we use the diagonal of the bootstrap covariance matrices to estimate the error bars on the auto-correlation function shown in Figure \ref{fig:measured_autocorr}. 
Specifically we define $\sigma_{\text{boot}} = \sqrt(\text{diag}(\boldsymbol{\Sigma_{\text{boot}}}))$.
The diagonal of the covariance matrix is not a full description of the error since the bins of the auto-correlation function are highly correlated and should thus fluctuate in a correlated way, thus making the full covariance matrix necessary in any computations. 
The error bars in Figure \ref{fig:measured_autocorr_norm}, $\sigma_{\Delta}$, come from combining the bootstrap estimate of the errors for $\xi_F$ with bootstrap estimate of the errors on $\langle F \rangle$ via:
\begin{equation}
    \sigma_{\Delta} = \frac{\xi_F - \langle F \rangle^2}{\langle F \rangle^2} \sqrt{\left(\frac{\sigma_{\text{boot}}}{\xi_F}\right)^2 + 2 \left(\frac{\sigma_F}{\langle F \rangle}\right)^2}
\end{equation}

\begin{figure*}
	\includegraphics[width=2\columnwidth]{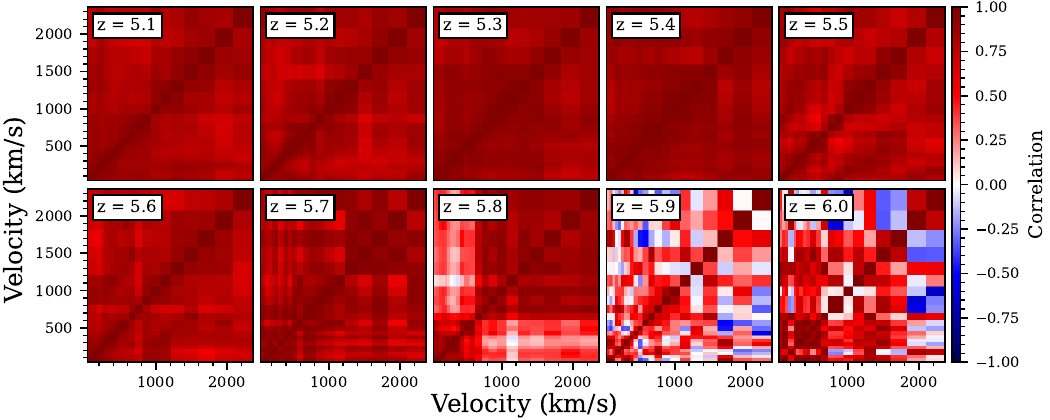}
    \caption{
        The correlation matrices from bootstrap re-sampling the auto-correlation function in the ten redshift bins considered in this work.
        For $z < 5.8$ we see very strong positive off-diagonal values of the correlation matrices.
        This behavior is expected since each pixel in the \lya forest contribute to every bin of the auto-correlation function, making these bins highly correlated. 
        The fluctuations in the correlation matrix values are caused by noise due to the limited sightlines available to bootstrap. 
        At $z \geq 5.8$ the number of sightlines is small and the transmission is low, causing large noise fluctuations. 
        For $z = 5.9$ and $z = 6.0$, the sightlines are so few and so non-transmissive that noise fluctuations lead to negative values in the correlation matrices. 
        There is no physical explanation for these negative values.
        The numbers of sightlines used at each $z$ are listed in Table \ref{table:quasar_los_flux}.
    }
    \label{fig:boot_covars}
\end{figure*}

Additionally we define the correlation matrix, $C$, which expresses the covariances between $j$th and $k$th bins in units of the the diagonal elements of the covariance matrix. 
This is done for the $j$th, $k$th element by
\begin{equation}
        C_{jk} = \frac{\Sigma_{jk}}{\sqrt{\Sigma_{jj}\Sigma_{kk}}}.
        \label{eq:correlation}
\end{equation}
The bootstrap correlation matrices for the measured auto-correlation functions at each $z$ are shown in Figure \ref{fig:boot_covars}. 
Based on the simulated correlation matrices from \citet{Wolfson_2022}, we expect there to be significant off diagonal values of these bootstrap correlation matrices. 
This is because, generally, each pixel in the \lya forest contributes to every bin of the auto-correlation function so the different velocity bins in the auto-correlation function are highly covariant. 
Large off-diagonal values are seen in the bootstrap correlation matrices in Figure \ref{fig:boot_covars} for $z < 5.8$. 
At the highest three redshifts, especially $z = 5.9$ and $z = 6.0$, the number of quasar lines of sight are quite small and the transmission is quite low, leading to large noise fluctuations and non-converged off-diagonal values. 
In particular, there are negative values off the diagonal for $z = 5.9$ and $z = 6.0$ which we do not see in our simulated covariance matrices. 
We expect noisy fluctuations in the off-diagonal covariance matrix values to go away with the addition of more quasar sightlines, though low transmission at the highest redshifts will still make this computation difficult.

\section{Modeling the measurement} \label{sec:sims}

In order to interpret the physical implications of the measured auto-correlation function, we construct forward models with the properties of the observed quasars. 
Functions to convert any set of simulation skewers into auto-correlation function measurements are available online.
In addition, there is a \texttt{Jupyter Notebook} that goes through an example of forward-modeling simulation skewers and then computing the auto-correlation function.  
The simulation method used here was introduced in \citet{Wolfson_2022} for a simplified mock data set. 
We have updated this method to include continuum uncertainty, noise vectors from observational data, and a $\guvb$ box that matches the density field of the main simulation suite. 
We will briefly describe this updated method here, for more information see \citet{Wolfson_2022}.

\subsection{Simulation box}

To begin, we use a \texttt{Nyx} simulation box \citep{almgren_2013}. 
\texttt{Nyx} is a hydrodynamical simulation code designed to simulate the Ly$\alpha$ forest with updated physical rates from \citet{lukic_2015}.
The \texttt{Nyx} box has a size of $L_{\text{box}} = 100$ cMpc $h^{-1}$ with $4096^3$ dark matter particles and $4096^3$ baryon grid cells. 
This box is reionized by a \citet{haardt_madau_2012} uniform UVB that is switched on at $z \sim 15$, which means these simulation boxes do not include the effects of a patchy, inhomogeneous reionization. 
An upcoming publication, Wolfson et al. in prep., investigates the effect of inhomogeneous reionization on the \lya forest flux auto-correlation function at $z = 5.8$ and found that it mainly affected scales $v < \SI{100}{\kilo\meter\per\second}$.
From this, we expect inhomogeneous reionization to at most affect our smallest bin so these models are sufficient for the comparison presented here. 
Additional work to explore the interactions between inhomogeneous reionization on a grid of $\lammfp$ values is left to future work. 

We have three snapshots of this simulation at $z = 5.0$, $z = 5.5$, and $z = 6$ and we want to model all ten redshifts $5.1 \leq z \leq 6.0$ with $\Delta z = 0.1$.
In order to consider the redshifts for which we do not have a simulation output, we select the nearest snapshot and use the desired redshift when calculating the proper size of the box and the mean density. 
This means we use the density fluctuations, temperature, and velocities directly from the nearest \texttt{Nyx} simulation output. 
Previously, in \citet{Wolfson_2022} they tested this choice of simulation interpolation by using the $z=6.0$ simulation snapshot to generate skewers at $z = 5.7$ and found no change in their results.

In addition, we have a grid of boxes of $\guvb/\langle \guvb \rangle$ values generated with the semi-numerical method of \citet{davies_furlanetto_2016} corresponding to a fluctuating UVB for different $\lammfp$ values, all at $z = 5.5$. 
These boxes have a size of $L_{\text{box}} = 100$ $h^{-1}$ cMpc, $64^3$ pixels, and are generated from the density field of the \texttt{Nyx} simulation box.
The method of \citet{davies_furlanetto_2016} uses \citet{mesinger_furlanetto_2007} and \citet{bouwens_2015} to create halos and assign UV luminosities from the density field.
They then get the ionizing luminosity of each galaxy by assuming it to be proportional to its UV luminosity where the constant of proportionality is left as a free parameter. 
Finally the ionizing background radiation intensity, $J_{\nu}$, is computed by a radiative transfer algorithm and $\guvb$ is finally calculated by integrating over  $J_{\nu}$.
For more information on this method of generating $\guvb$ boxes see \citet{davies_furlanetto_2016}, \citet{davies_2018_abc}, or \citet{Wolfson_2022}. 

To combine the \texttt{Nyx} box with the $\guvb$ values generated via the \citet{davies_furlanetto_2016} method, we linearly interpolated $\log(\guvb/\langle \guvb \rangle)$ onto the higher resolution grid of the \texttt{Nyx} simulation box. 
We then re-scale the optical depths from the \texttt{Nyx} box with a constant UVB, $\tau_{\text{const.}}$, by these fluctuating $\guvb$ values to get the optical depths for a fluctuating UVB, $\tau_{\text{mfp}} = \tau_{\text{const.}}/(\guvb/\langle \guvb \rangle)$. 
This implies that we need to know $\langle \guvb \rangle$ to compute our final optical depths, which is not known a priori. 
We therefore determine this value by matching an overall mean flux $\langle F \rangle$, where we vary $\langle F \rangle$ over a range of models based off the measurement of \citet{bosman_2021_data}. 
We generate 1000 skewers from this simulation method for each $\lammfp$ and $\langle F \rangle$ at each $z$ for $5.1 \leq z \leq 6.0$. 
These skewers come from the same location in the simulation box for all parameter values and $z$.

\subsection{Forward modeling} \label{sec:forward_modeling}

Our simulations provide skewers of the optical depth of the \lya forest for given $\lammfp$ and $\langle F \rangle$ values.
In order to compare these (or any) simulated skewers to the results of our observational measurement, we forward model the telescope resolution, the noise properties of our observed sightlines, and the continuum uncertainty from the PCA continuum fit. 
This section will describe how each property is modeled for our simulation skewers, see the \texttt{lya-autocorr} git repository to follow along with an example simulation skewers being forward modeled.

To model the resolution of  X-shooter for visible light with a 0.9" slit, we convolved the flux by a Gaussian line-spread function with $\text{FWHM} \approx \SI{34}{\kilo\meter\per\second}$.
This corresponds the nominal resolving power ($R \sim 8800$) of the X-Shooter setup used for the XQR-30 data. 
However, as noted in Section \ref{sec:data} the actual data has a higher median resolving power in the visible of $R = 11400$ \citep{dodorico_2023_xqr30}. 
Future work will use the measured resolving power for each quasar in the modeling but using the nominal value for all is sufficient for this initial comparison. 
After using this Gaussian filter we interpolated the line-spread-function convolved flux onto the exact velocity grid from the observation. 
This step also reduced the simulation skewers from the box size to the same length as our observations, as 100 cMpc $h^{-1}$ corresponds to $\Delta z \sim 0.3$ at the relevant redshifts and our observations have $\Delta z = 0.1$.

We add noise to the interpolated, line-spread-function convolved flux, $\boldsymbol{F_{\text{res}}}$, according to the noise vectors for each quasar sightline, $\boldsymbol{\sigma_{\text{qso}}}$, with random normal distribution realization, $\boldsymbol{N_{\text{qso}}} \sim N(0, 1)$, via
\begin{equation}
    \boldsymbol{F_{\text{noise}}} = \boldsymbol{F_{\text{res}}} + \left( \boldsymbol{N_{\text{qso}}} \times \boldsymbol{\sigma_{\text{qso}}} \right).
\end{equation}
$\boldsymbol{F_{\text{noise}}}$ is thus the flux modeled with both the resolution of the telescope and the noise properties of our observed sightlines. 

To model continuum error, we used the mean, $\boldsymbol{\mu_{\text{cont}}}$, and covariance, $\boldsymbol{\Sigma_{\text{cont}}}$, of the PCA reconstruction just as we do for the data as described in Section \ref{sec:continuum fitting}.
We randomly draw realizations of the continuum error, $\boldsymbol{E_{\text{cont}}} \sim N(\boldsymbol{\mu_{\text{cont}}}, \boldsymbol{\Sigma_{\text{cont}}})$, where $N$ is the normal distribution. 
% We can then forward model the predicted continuum with error, $\boldsymbol{C_{\text{pred}}}$, from the fit quasar continuum, $\boldsymbol{C_{\text{fit}}}$, via: 
% \begin{equation}
%     \boldsymbol{C_{\text{pred}}} = \boldsymbol{C_{\text{fit}}} \times \boldsymbol{E_{\text{cont}}}.
% \end{equation}
In our simulations we do not fit and normalize by the quasar continuum so we model continuum error by:
\begin{equation}
    \boldsymbol{F_{\text{cont}}} = \boldsymbol{F_{\text{noise}}} / \boldsymbol{E_{\text{cont}}}
\end{equation}
where $\boldsymbol{F_{\text{cont}}}$ is the final fully forward-modeled \lya forest spectra.  
We investigate the effect of the continuum modeling on the resulting models of the auto-correlation function in Appendix \ref{appendix:continuum}. 

Ultimately, we generate $N_{\text{skewer}}$ forward-modeled copies of each of the $N_{\text{los}}$ quasars in the sample, where $N_{\text{skewer}} = 1000$ from the simulation and $N_{\text{los}}$ is the number of quasar sightlines at each redshift as listed in Table \ref{table:quasar_los_flux}.
% Ultimately, we forward model each simulation skewers for all $N_{\text{los}}$ observed quasars, where $N_{\text{los}}$ is the number of quasar sightlines at each redshift as listed in Table \ref{table:quasar_los_flux}, creating $N_{\text{los}}$ copies of each simulation skewer. 
For example at $z = 5.1$ we have $N_{\text{skewers}} \times N_{\text{los}} = 1000 \times 24 = 24000$ total forward-modeled \lya forest spectra.

\begin{figure*}
	\includegraphics[width=2\columnwidth]{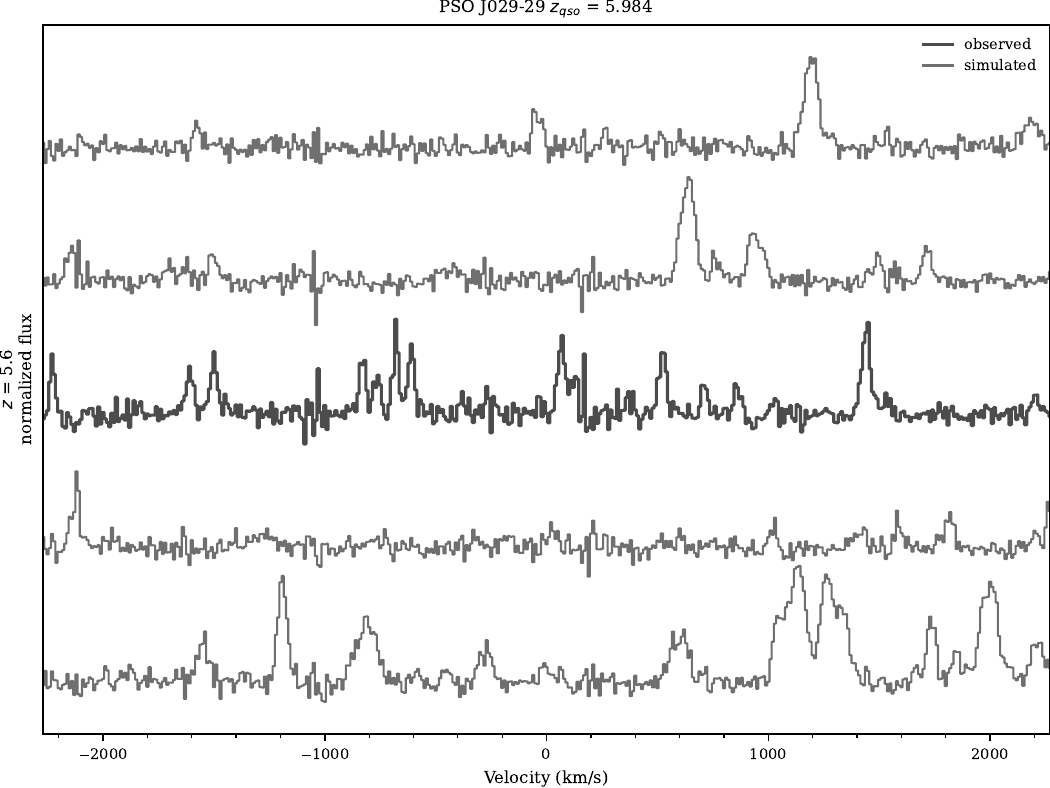}
    \caption{
        This figure compares the observed \lya forest flux at $z = 5.6$ from PSO J029-29 with forward modeled simulation skewers modeled to have the same noise properties as this quasar. 
        The thick line in the middle is the observed flux while the other four thinner lines are from the forward modeled simulations. 
        The visual similarities between the observed and simulated \lya forest flux shown here demonstrates the success of our forward-modeling procedure. 
    }
    \label{fig:forward_flux}
\end{figure*}

Figure \ref{fig:forward_flux} shows the normalized flux of the $z = 5.6$ \lya forest from PSO J029+29 with four examples of the normalized flux from our simulations that were forward modeled with this quasar's properties. 
The thick line in the middle is the flux from the quasar while the other four thinner lines are from the simulation.
The visual similarities between the observed data and the forward modeled data highlights the ability of our forward modeling methods to mimic realistic data. 
The remaining figures all show data and simulations at $z = 5.6$ because this redshift has the maximal observed sightlines with $N_{\text{los}} = 34$ and there is a nearby measurement of $\lammfp$ at $z = 5.6$ by \citet{zhu_2023}.
Note that $N_{\text{los}}$ does not affect the convergence of our simulated models but it determines the convergence of the bootstrap covariance matrix estimate which we will compare to later in the section.

\subsection{Modeled auto-correlation function}

We then computed the auto-correlation function of these forward modeled skewers in the same way as the actual data, with equation \eqref{eq:autocorr}, for each copy of the skewer.
We used the same mask from the observed quasar when computing the auto-correlation function. 
This includes the DLA mask as described in Section \ref{sec:dla_mask}.
In the observations, the DLA mask corresponds to regions in the spectra where the transmission is low. 
However, for the simulations the DLA mask corresponds to random parts of the spectra.
We choose to include this part of the mask for the simulation data in order to keep the number of pixel pairs used per quasar sightline the same between simulations and observations. 
A discussion on the effect of the DLA mask on the measured auto-correlation function can be found in Appendix \ref{appendix:dla}. 

To create a mock data set, we randomly selected $N_{\text{los}}$ quasars from the 1000 forward modeled skewers without replacement. 
We then assigned each of the randomly selected skewers one of each of the $N_{\text{los}}$ quasars, so each mock data set had exactly one skewer forward modeled with the properties of copy each quasar.  
The value of the auto-correlation function from the mock data set, $\xi_{i}$, is then the weighted average of the auto-correlation function from these $N_{\text{los}}$ forward modeled skewers, where the weights are the number of pixels pairs in each bin of the auto-correlation function. 
We defined the model value of the auto-correlation function, $\boldsymbol{\xi}_{\text{model}} = \boldsymbol{\xi}_{\text{model}}(\lammfp, \langle F \rangle)$, to be the weighted average of the auto-correlation functions from all $N_{\text{los}} \times N_{\text{skewers}}$ skewers generated.
The simulated covariance matrices, $\boldsymbol{\Sigma_{\text{sim}}}$, are computed for each $\lammfp$ and $\langle F \rangle$ values from $N_{\text{mocks}}$ mock data sets in the following way:
\begin{equation}
    \boldsymbol{\Sigma_{\text{sim}}}(\boldsymbol{\xi_\text{model}}) = \frac{1}{N_{\text{mocks}}} \sum_{i=1}^{N_{\text{mocks}}}(\boldsymbol{\xi}_i - \boldsymbol{\xi_\text{model}})(\boldsymbol{\xi}_i - \boldsymbol{\xi_\text{model}})^{\text{T}}.
    \label{eq:covariance}
\end{equation}

\begin{figure*}
	\includegraphics[width=2\columnwidth]{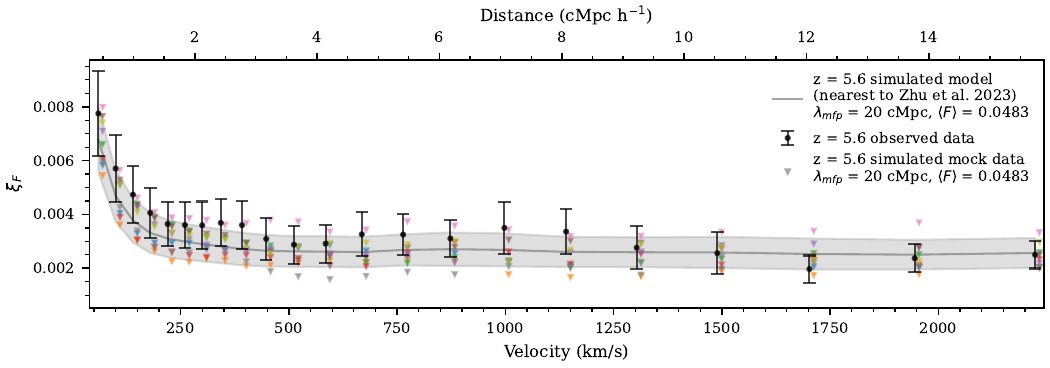}
    \caption{
        The black points show the observed auto-correlation function from the extended XQR-30 data discussed in this work at $z = 5.6$. 
        The colored triangles show the auto-correlation value for 9 different simulated mock data sets. 
        The mock data sets shown here were all modeled with $\lammfp = 20$ cMpc and $\langle F \rangle = 0.0483$, the closest $\lammfp$ value to the \citet{zhu_2023} measurement and the closest $\langle F \rangle$ value to our measurement listed in Table \ref{table:quasar_los_flux}. 
        The model value of the auto-correlation is shown as the grey line with the shaded region representing the diagonal elements from the corresponding simulated covariance matrix. 
    }
    \label{fig:mock_auto_corr}
\end{figure*}

Figure \ref{fig:mock_auto_corr} shows nine measurements of the auto-correlation function from nine different mock data sets generated from the simulations at $z = 5.6$ (colored triangles). 
These mock measurements were generated from the $\lammfp = 20$ cMpc and $\langle F \rangle = 0.0483$ simulation, the closest $\lammfp$ value to the \citet{zhu_2023} measurement and the closest $\langle F \rangle$ value to our measurement listed in Table \ref{table:quasar_los_flux}. 
This model value of the auto-correlation function is shown as the grey line where the grey shaded region shows the error from the diagonal of the simulated covariance matrix.
The black points show the measured auto-correlation function at $z = 5.6$ with error bars from the bootstrap covariance matrix. 
This plot demonstrates that our forward modeling procedure leads to mock correlation function measurements that are visually similar to our actual measurement.
This plot also shows that our measured auto-correlation function and the model with the value from \citet{zhu_2023} agree within $1 \sigma_{\text{boot}}$ for nearly all the points, though again these errors come from the diagonal of the covariance matrix only and therefore do not include information on the strong off-diagonal covariance between auto-correlation function bins. 
We discuss the comparison of our measured auto-correlation function and the measurements of \citet{zhu_2023} and \citet{gaikwad_2023_mfp} in Section \ref{sec:compare}.

\subsection{Model based covariance matrices}

\begin{figure}
	\includegraphics[width=\columnwidth]{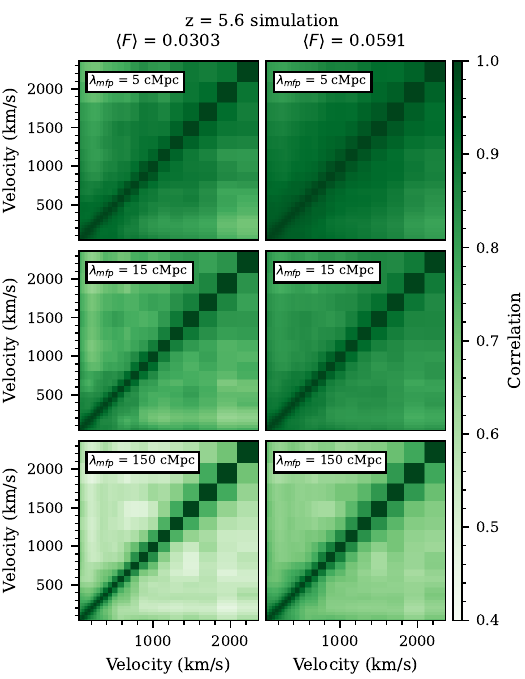}
    \caption{
        Correlation matrices for six different simulation model values at $z = 5.6$. 
        These six covariance matrices come from the combination of $\lammfp = 5, 15, 150$ cMpc and $\langle F \rangle = 0.0303, 0.0591$ as labeled in the title of each subplot. 
        These include the maximal and minimal $\lammfp$ and $\langle F \rangle$ values simulated at $z = 5.6$. 
        This shows the model-dependence of the correlation (and thus covariance) matrices. 
        Larger values of $\lammfp$ result in weaker off-diagonal correlation matrix values, as is seen going down the rows. 
        Smaller $\langle F \rangle$ values also appear to cause weaker off-diagonal correlation matrix values (as seen when comparing the left and right columns) but this effect is weaker than the effect of $\lammfp$. 
    }
    \label{fig:forward_covars}
\end{figure}

Figure \ref{fig:forward_covars} shows correlation matrices from the forward modeled data for six different parameter values at $z = 5.6$. 
The parameter values shown are $\lammfp = 5, 15, 150$ cMpc going down the rows and then $\langle F \rangle = 0.0303, 0.0591$ across the columns, both of which span the full range of parameter values available to us. 
Going from the left to the right column, we see that increasing the $\langle F \rangle$ weakly increases the off-diagonal values of the correlation matrices, however the effect going down the rows is much stronger. 
Going down the rows shows that an increase in $\lammfp$ decreases the off-diagonal values for the correlation matrix.
This means that shorter $\lammfp$ models have more highly covariant bins in the auto-correlation function.

\begin{figure}
	\includegraphics[width=\columnwidth]{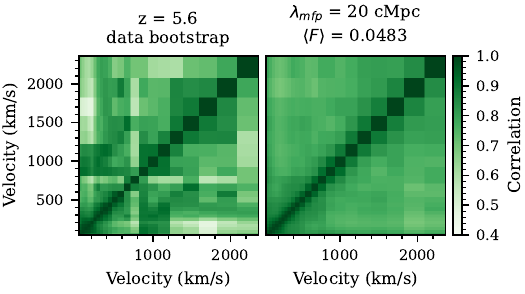}
    \caption{
        The correlation matrix computed via bootstrap re-sampling the data at $z = 5.6$ (left) and the simulated correlation matrix from the model with $\lammfp = 20$ cMpc and $\langle F \rangle = 0.0483$ (right). 
        This model was chosen as the model with the closest $\lammfp$ value to the \citet{zhu_2023} measurement and the closest $\langle F \rangle$ value to our measurement.
        This bootstrap correlation matrix is also shown in Figure \ref{fig:boot_covars} with a different color bar and has been reproduced here with the color bar used in Figure \ref{fig:forward_covars}, to more easily compare the values of the correlation matrix from data to the simulated examples. 
        The bootstrap covariance matrix is noisy due to the limited data available, though this redshift was selected as the bin with the maximal value of $N_{\text{los}} = 34$.
    }
    \label{fig:one_bootstrap_z56}
\end{figure}

To compare a bootstrap covariance matrix from the data with the forward modeled covariance matrices, Figure \ref{fig:one_bootstrap_z56} shows the bootstrap correlation matrix at $z = 5.6$ with the same color bar as Figure \ref{fig:boot_covars}.
Additionally, Figure \ref{fig:one_bootstrap_z56} shows the simulated correlation matrix for the $\lammfp = 20$ cMpc and $\langle F \rangle = 0.0483$ model to directly compare to the bootstrapped matrix. 
Again, this is the model with the closest $\lammfp$ value to the \citet{zhu_2023} measurement and the closest $\langle F \rangle$ value to our measurement.
The bootstrap covariance matrix is still quite noisy due to the limited data available so it is difficult to determine the best matching simulated covariance matrix.
The bootstrap correlation matrix has regions of high off diagonal values, such as $\SI{1200}{\kilo\meter\per\second} < v < \SI{2000}{\kilo\meter\per\second}$ as well as individual pixels with relatively small off-diagonal values, such as the combination of $v = \SI{60}{\kilo\meter\per\second}$ and $v = \SI{1702}{\kilo\meter\per\second}$. 
This potentially suggests additional structure in the bootstrap covariance matrix compared to the simulated covariance data, but these fluctuations appear consistent with the noise.

As can be seen in Figure \ref{fig:forward_covars}, the correlation matrices, and therefore the covariance matrices, strongly depend on the model value of $\lammfp$. 
For this reason, when attempting to fit this data to a model, we would be fitting both the measured auto-correlation function as well as the covariance structure between the bins. 
While the amplitude of the correlation function might favor one combination of model parameters, it is conceivable that the level of fluctuations between two correlated correlation function bins, which is quantified by the covariance matrix, could favor a different combination of parameters. 
For this reason, fitting these models to our measurements is quite challenging and we leave this discussion for future work.

\subsection{Comparison to previous work} \label{sec:compare}

\begin{figure*}
	\includegraphics[width=2\columnwidth]{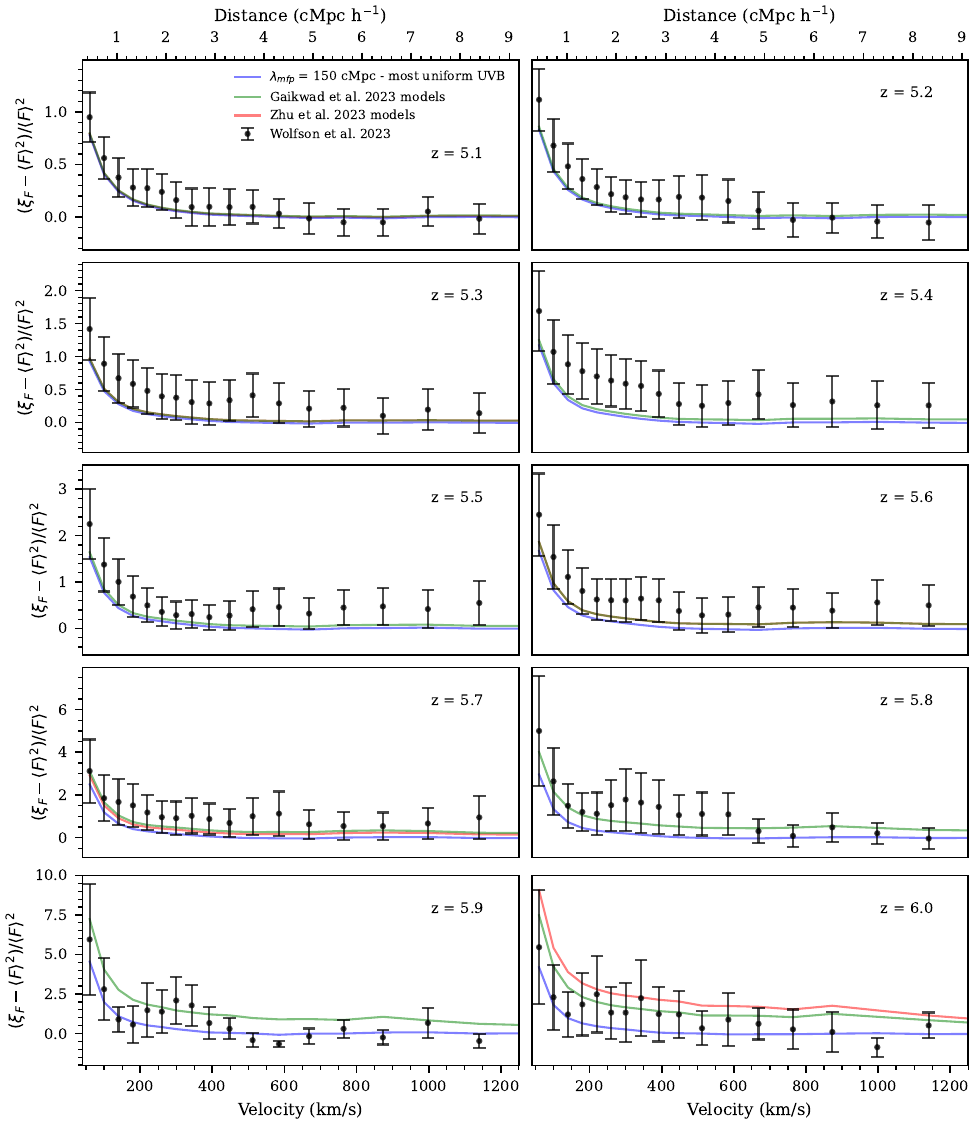}
    \caption{
        The auto-correlation function of \lya transmission normalized and shifted by the mean transmission, $\langle F \rangle$, in ten redshift bins measured in this work.
        \citet{gaikwad_2023_mfp} measured $\lammfp$ at each of these redshifts and so each panel has our model with their $\lammfp$ values (green lines). 
        \citet{zhu_2023} has measured $\lammfp$ for $z =$ 5.08, 5.31, 5.65, and 5.93. 
        We show the model models for the measured $\lammfp$ values from \citet{zhu_2023} in the $z =$ 5.1, 5.3, (5.6 and 5.7), and 6.0 panels respectively (red lines). 
        The model for a uniform UVB value (blue line) is also shown as a comparison. 
        }
    \label{fig:compare_measured_autocorr}
\end{figure*}

We model the auto-correlation function at any value of $\lammfp$ and $\langle F \rangle$ via nearest grid-point emulation from our initial grid of values.
Therefore, we can compare our auto-correlation function measurement to the models with the $\lammfp$ values measured in \citet{gaikwad_2023_mfp} and \citet{zhu_2023} which updated the measurements of \citet{becker_2021}. 
Since we need to specify both $\lammfp$ and $\langle F \rangle$ to get our models, we use the measured $\langle F \rangle$ from this work to get the models representing the $\lammfp$ values from the corresponding alternative measurements. 
Figure \ref{fig:compare_measured_autocorr} has ten panels, each of which has one of our measured $\xidif$ values (shown as the black points) at a given $z$. 
We have chosen to show $\xidif$ instead of the regular auto-correlation function because we have to use the nearest grid point on a coarse $\langle F \rangle$ grid which could be quite far from the measured $\langle F \rangle$ value. 
This would have a large effect on the auto-correlation function value and a smaller effect on $\xidif$. 

\citet{gaikwad_2023_mfp} measured $\lammfp$ at each of these redshifts and so each panel has our model with their $\lammfp$ values (green lines). 
\citet{zhu_2023} has measured $\lammfp$ for $z =$ 5.08, 5.31, 5.65, and 5.93. 
We show the models for the measured $\lammfp$ values from \citet{zhu_2023} in the $z =$ 5.1, 5.3, (5.6 and 5.7), and 6.0 panels respectively (red lines). 
Finally, we also show the model for $\lammfp = 150$ cMpc, our most uniform UVB (blue line). 

Making a quantitative comparison of these models with the measured auto-correlation function is difficult due to the expected large-off diagonal values of the covariance matrix as well as the noise in the bootstrap covariance matrices as shown in Figure \ref{fig:boot_covars}. 
For this reason we leave detailed quantitative comparisons and fitting for future work. 
It is interesting to note that our measurements fall above the models from \citet{zhu_2023}, \citet{gaikwad_2023_mfp}, and $\lammfp = 150$ cMpc for $z < 5.8$.
Also note that models from \citet{zhu_2023} and \citet{gaikwad_2023_mfp} show a small boost over the most uniform UVB model for $z < 5.8$. 

\section{Conclusions} \label{sec:conclusion}

In this work we have measured the auto-correlation function of the \lya forest flux from the extended XQR-30 data set in 10 redshift bins, $5.1 \leq z \leq 6.0$.
This is the first measurement of the auto-correlation function of the \lya forest at these redshifts. 
Our final assembled data set includes 36 $z > 5.7$ quasars with SNR $> 20$ per spectral pixel.
This data set was analyzed while fully accounting for the error from continuum reconstruction, instrumentation, and contamination from DLAs. 
We measured the average transmission, $\langle F \rangle$, from this data and found good agreement with previous work. 
We found that the boost in the auto-correlation function on the smallest scales increases when increasing $z$, which may suggest a decrease in $\lammfp$.
We additionally measured covariance matrices of the auto-correlation function by bootstrap re-sampling the available data. 
The convergence of these matrices was hindered by noise from the limited number of sightlines and low transmission, especially for the highest redshift bins, $z \geq 5.8$. 
The auto-correlation function measurements as well as the bootstrap covariance matrices are available to download online at \texttt{https://github.com/mollywolfson/lya\_autocorr/}.

In addition, we introduced \lya forest simulations with a fluctuating UVB model described by $\lammfp$. 
This comparison indicates preliminary agreement between these models and our measurements. 
We found that the covariance matrices produced from the simulations had a strong dependence on $\lammfp$. 
In order to fit these models to our data, we would need to use an estimate of the covariance matrix for the bins of the auto-correlation function. 
In this work we have presented two options for this covariance matrix: the bootstrap estimate, $\boldsymbol{\Sigma_{\text{boot}}}$, and the simulation covariance matrices, $\boldsymbol{\Sigma_{\text{sim}}}$. 
Ideally we would like to use $\boldsymbol{\Sigma_{\text{boot}}}$ when fitting, however as seen in Figure \ref{fig:boot_covars}, these covariance matrices are quite noisy and non-converged. 
Therefore, we could hope to use $\boldsymbol{\Sigma_{\text{sim}}}$, where the off-diagonal structure depends strongly on the value of $\lammfp$. 
This dependence of $\boldsymbol{\Sigma_{\text{sim}}}$ on $\lammfp$ means that fitting the models to the data would require fitting both the mean line as well as this covariance structure, which is subtle. 
Thus, additional work is necessary to get robust measurements of $\lammfp$, which we leave to the future.
We did show a preliminary comparison of our measured auto-correlation function to models with the $\lammfp$ values measured by \citet{gaikwad_2023_mfp} and \citet{zhu_2023}, leaving a quantitative comparison of these results to future work. 

With this work we have included a link to a Git repository with the code necessary to measure the auto-correlation function from any set of simulation skewers. 
This will allow other simulation groups to compare the auto-correlation function from their simulations to our measured auto-correlation function and thus foster more work on this statistic.

Future work to get a robust measurement of $\lammfp$ from the \lya forest auto-correlation function include further considerations in the modeling methods. 
The \citet{davies_furlanetto_2016} method to generate $\guvb$ for various $\lammfp$ assumes a fixed source model.
Other source model choices could impact the fluctuations in $\guvb$ seen at a fixed $\lammfp$ value, and thus bias measurements from observation data when compared with these models. 
Additionally, rare bright sources could cause boosts in the auto-correlation function for individual sightlines that aren't modeled in our simulations. 
We leave a detailed investigation into these effects on the auto-correlation function models and covariance matrices to future work. 

Note that in order to generate UVB fluctuations due to different $\lammfp$ values that matched the density field of our \texttt{Nyx} simulation, we also generated UVB fluctuations in a 100 cMpc h$^{-1}$ box. 
\citet{Wolfson_2022} found that using a 40 cMpc h$^{-1}$ box to generate UVB fluctuations significantly reduced the auto-correlation function on all scales when compared to a 512 cMpc box. 
Future work would be needed to understand the effect of the box size on any measured $\lammfp$ from the auto-correlation function with a 100 cMpc h$^{-1}$ UVB box. 

Additionally, this work ignored the effect of inhomogeneous reionization beyond a fluctuating UVB. 
It is expected that a patchy, inhomogeneous reionization process would have other physical effects, such as additional fluctuations in the thermal state of the IGM. 
We leave an exploration of the effect of the temperature of the IGM on the \lya forest flux auto-correlation function, including the effect of temperature fluctuations, to a future work.

Overall, this first measurement of the $z > 5$ \lya forest flux auto-correlation functions opens up an exciting new way to measure $\lammfp$ at the tail-end of reionization.

\section*{Acknowledgements}
We acknowledge helpful conversations with the entire XQR-30 collaboration as well as the ENIGMA group at UC Santa Barbara and Leiden University. 
JFH acknowledges support from the European Research Council (ERC) under the European Union’s Horizon 2020 research and innovation program (grant agreement No 885301) and from the National Science Foundation under Grant No. 1816006.
MH acknowledges support by STFC (grant number ST/N000927/1).
GK is partly supported by the Department of Atomic Energy (Government of India) research project with Project Identification Number RTI~4002, and by the Max Planck Society through a Max Planck Partner Group.

Based on observations collected at the European 761 Southern Observatory under ESO programme 1103.A0817.
This research was supported in part by the National Science Foundation under Grant No. NSF PHY-1748958.
This research used resources of the National Energy Research Scientific Computing Center (NERSC), a U.S. Department of Energy Office of Science User Facility located at Lawrence Berkeley National Laboratory, operated under Contract No. DE-AC02-05CH11231.

For the purpose of open access, the authors have applied a Creative Commons Attribution (CC BY) licence to any Author Accepted Manuscript version arising from this submission.

%%%%%%%%%%%%%%%%%%%%%%%%%%%%%%%%%%%%%%%%%%%%%%%%%%
\section*{Data Availability}

The XQR-30 spectra and associated meta-data have been made public in \citet{dodorico_2023_xqr30}.
The measurements of the auto-correlation function are available to download online at \texttt{https://github.com/mollywolfson/lya\_autocorr/} along with functions to forward model any simulation skewers with the properties of our data set. 

The simulation data analyzed in this article will be shared on reasonable request to the corresponding author.

% The inclusion of a Data Availability Statement is a requirement for articles published in MNRAS. Data Availability Statements provide a standardised format for readers to understand the availability of data underlying the research results described in the article. The statement may refer to original data generated in the course of the study or to third-party data analysed in the article. The statement should describe and provide means of access, where possible, by linking to the data or providing the required accession numbers for the relevant databases or DOIs.



\bibliographystyle{mnras}
\bibliography{xqr30_measure_1} % if your bibtex file is called example.bib

%\begin{thebibliography}{99}
%\end{thebibliography}

%%%%%%%%%%%%%%%%%%%%%%%%%%%%%%%%%%%%%%%%%%%%%%%%%%

%%%%%%%%%%%%%%%%% APPENDICES %%%%%%%%%%%%%%%%%%%%%

\appendix

\section{Continuum uncertainty modeling effect} \label{appendix:continuum}

\begin{figure*}
	\includegraphics[width=2\columnwidth]{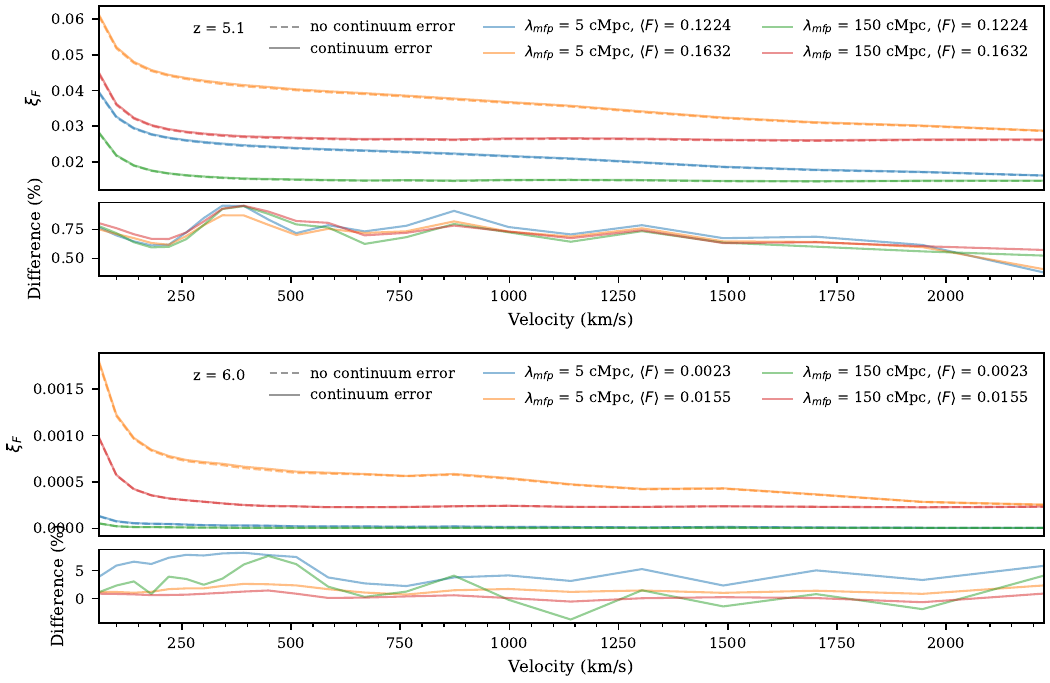}
    \caption{
        The first and third panels show the auto-correlation function from the simulations with (solid line) and without (dashed line) modeling continuum uncertainty at redshifts of 5.1 and 6.
        The different colors represent different parameter values of $\lammfp$ and $\langle F \rangle$ used. 
        The second and fourth panels show the relative difference between these lines defined by Equation \eqref{eq:continuum_dif}, in percent. 
    }
    \label{fig:app_continuum}
\end{figure*}

Figure \ref{fig:app_continuum} quantifies the difference in the auto-correlation models calculated from forward-modeled skewers with or without continuum uncertainty multiplied in, as described in Section \ref{sec:forward_modeling}.
The first and third panels show the auto-correlation function from the simulations with (solid line) and without (dashed line) modeling continuum uncertainty at $z = 5.1$ and 6.
The different colors represent different parameter values of $\lammfp$ and $\langle F \rangle$ used. 
The second and fourth panels show the relative difference in percent, defined as:
\begin{equation}
    \frac{\xi_{\text{cont}} - \xi_{\text{no cont}}}{\xi_{\text{no cont}}}.
    \label{eq:continuum_dif}
\end{equation}

At $z = 5.1$ there is a $<1\%$ of a difference between the auto-correlation models with and without the continuum error. 
At $z = 6.0$ there is a larger difference between the models where the difference is $<8\%$ for all the parameter values. However, the effect is most noticeable when $\langle F \rangle$, and hence the auto-correlation function which goes as $\langle F \rangle^2$, is quite small. 
For the other $\langle F \rangle$ value at this redshift the error is $<2\%$. 
These values are typically positive because of the bias in the continuum reconstruction as seen in Figure 2 of \citet{bosman_2021_data}. 

We computed the difference in our measured auto-correlation function at all $z$ with and without continuum error. 
The difference in the measured data ranges from at most 0.4\% to 1.8\%  with a stronger effect at the highest redshifts.

\section{DLA modeling effect} \label{appendix:dla}

\begin{figure*}
	\includegraphics[width=2\columnwidth]{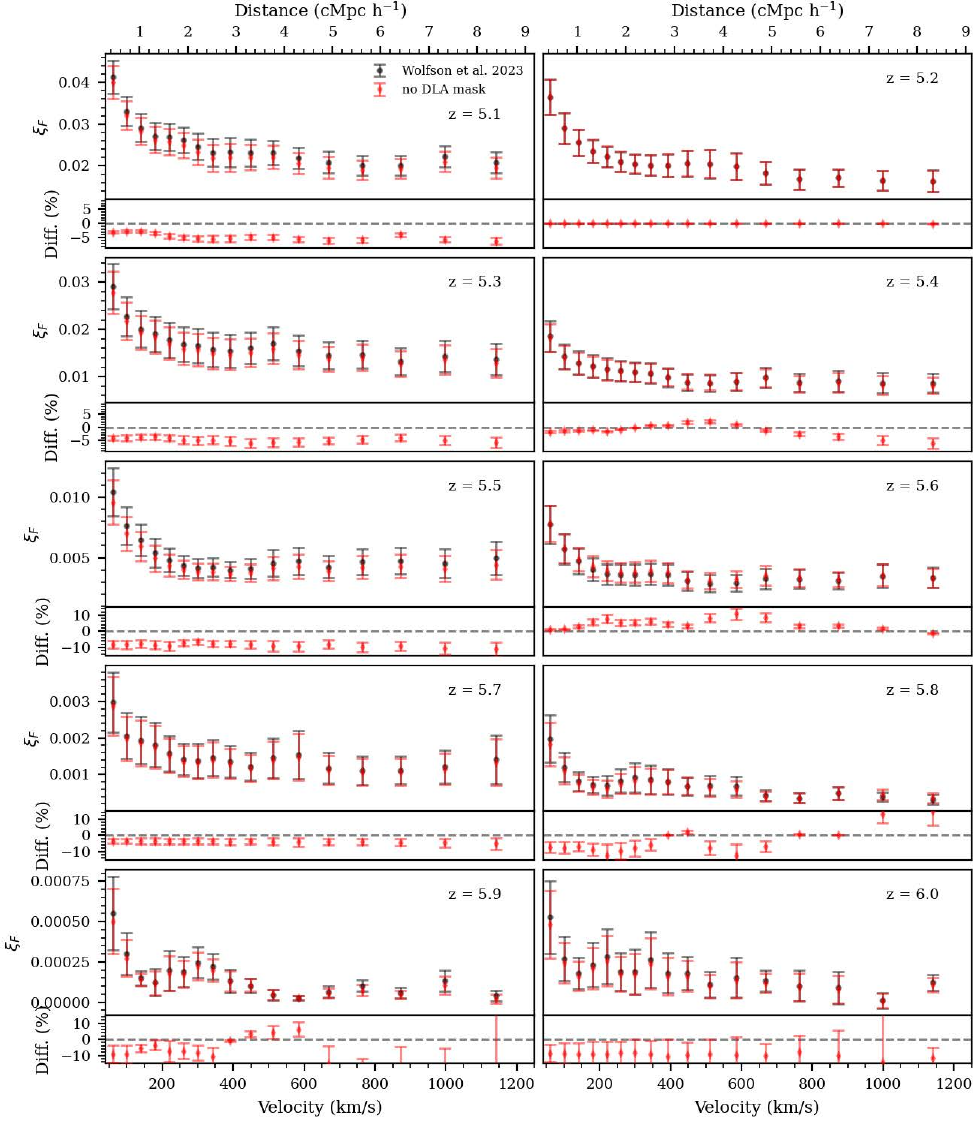}
    \caption{
        The black points show auto-correlation function of \lya transmission in ten redshift bins measured in this work. 
        The red points show the measured auto-correlation function of the \lya transmission when ignoring the masks for the DLAs as described in Section \ref{sec:dla_mask}.
        In general, ignoring the DLA mask decreases the auto-correlation function values. 
    }
    \label{fig:app_dlas}
\end{figure*}

In order to investigate how the DLA mask that was described in Section \ref{sec:dla_mask} we compute the measured auto-correlation functions without using this mask. 
This is shown for all redshifts in Figure \ref{fig:app_dlas} in red. 
The original measurement including this mask is shown in black. 
The measurement at $z = 5.2$ is not impacted at all by the DLA mask as no sightline has a detected DLA in this redshift range. 
Otherwise, for most scales at most redshifts ignoring the DLA mask reduces the auto-correlation function values. 
This follows as generally the regions masked in our procedure are regions with high absorption.

% If you want to present additional material which would interrupt the flow of the main paper,
% it can be placed in an Appendix which appears after the list of references.

%%%%%%%%%%%%%%%%%%%%%%%%%%%%%%%%%%%%%%%%%%%%%%%%%%

% Don't change these lines
\bsp	% typesetting comment
\label{lastpage}
\end{document}